\documentclass[14pt]{extarticle}

\usepackage{amssymb,amsmath,amscd}
\usepackage{bm}
\usepackage{subfigure}
\usepackage{multicol}

\usepackage{mathtext}         
\usepackage[makeroom]{cancel} 
\usepackage[utf8]{inputenc}  

\usepackage{amssymb}                                          
\usepackage{amsmath}                                          
\usepackage{amssymb}

\usepackage{graphics}
\usepackage{graphicx}
\usepackage{changes}

\graphicspath{{pictures/}}
\usepackage{xcolor}
\usepackage{wrapfig,epsfig,epsf}
\usepackage{bm}    
\usepackage{hyperref}
\graphicspath{{pictures/}}

\makeatletter
\renewcommand\@biblabel[1]{#1.} 
\makeatother

\def\sovast{Sov. Astron.}

\def\prd{Phys. Rev. D}
\def\aj{AJ}
\def\apj{ApJ}
\def\apjl{ApJL}
\def\apjs{ApJ Suppl. Ser.}

\def\apspr{Astroph. Sp. Phys. Rev.}
\def\aap{A\&A}

\def\araa{ARAA}
\def\mnras{MNRAS}

\def\nat{Nature}
\def\na{New Astron.}
\def\nar{New Astron. Rev.}
\def\pasj{PASJ}

\def\skytel{Sky \& Telescope}

\usepackage[left=2.5cm,right=2.5cm,top=4cm,bottom=2.7cm]{geometry}

\def\beq#1{\begin{equation}\label{#1}}
\def\eeq{\end{equation}}
\def\beqa#1{\begin{eqnarray}\label{#1}}
\def\eeqa{\end{eqnarray}}

\def\myfrac#1#2{\left(\frac{#1}{#2}\right)}

\def\comment#1{\relax}

\newcommand{\dpb}{\dot P_\mathrm{orb}}
\newcommand{\mv}{M_\mathrm{v}}
\newcommand{\mx}{M_\mathrm{x}}
\newcommand{\dmv}{\dot M_\mathrm{v}}
\newcommand{\dmx}{\dot M_\mathrm{x}}

\title{Unique microquasar SS433: new results, new issues}
\author{
A.M. Cherepashchuk$^1$,\\
A.V. Dodin$^{1}$,\\
K.A. Postnov$^{1,2}$
\\
$^1$Sternberg Astronomical Institute, 13 Universitetskij pr., 119234 Moscow, Russia\\
$^2$Kazan Federal University, 18 Kremlyovskaya st., Kazan, Russia\\
}

\begin{document}

\maketitle

\tableofcontents

\begin{abstract}\
The unique microquasar SS433 is a massive X-ray binary system at an advanced stage of evolution. The
optical star overflows its Roche lobe and transfers mass at a very strong rate onto a black hole, around
which a supercritical accretion disk inclined to the orbital plane has formed with relativistic collimated
outflows (jets). Both disk and jets precess with a period of 162.3 days. In the outer parts of the
precessing jets, emission lines of hydrogen and neutral helium are formed, which move periodically
across the spectrum of SS433 with an enormous amplitude of $\sim 1000$\,\AA\ or, on the $\sim
50000$ km/s velocity scale. This unique feature of SS433 attracted much attention of scientists in 1979.
Over many years of research in the optical, infrared, radio, X-ray and gamma-ray ranges, many
important results have been obtained about the physical processes occurring in this microquasar, but a
number of fundamental questions about the nature of SS433 remained unresolved.

A 30-year spectral and photometric monitoring of SS433 has been carried out at Sternberg Astronomical
Institute of Moscow University. Using all published data for 45 years of observations, we obtained a
number of important results concerning the nature of this unique microquasar.
We discovered a secular evolutionary increase in the orbital period of SS433 at a rate of $(1.14 \pm
0.25)\times 10^{-7}$ seconds per second. On this basis, it is shown that the relativistic object in SS433 is
a black hole with mass exceeding 8\,${\rm M}_\odot$. It is shown that the distance between the
components of SS433 increases with time, which prevents the formation of a common envelope in the
system. The size of the Roche lobe of the optical donor star is on average constant in time, which
ensures a stable secondary mass exchange in the system

The orbital ellipticity of SS433 was discovered, strongly supporting the model of a slaved accretion disk
tracking the precession of the rotation axis of the optical star, which is inclined to the orbital plane due
to an asymmetric supernova explosion accompanying the formation of a relativistic object.
The parameters of the kinematic model of the system, except for the precession period, keep on
average constant for 45 years. Phase shifts of the precession period were detected, but on average the
precession period remains constant for 45 years.

Microquasar SS433 is physically similar to many ultra-luminous X-ray sources (ULX) discovered in recent
years in other galaxies. The registration of hard gamma-ray emission up to 200 TeV from the W50
nebula indicates a possible acceleration to $\sim$ PeV hadrons in the region of the interaction between
the powerful equatorial wind from SS433 and the matter of the nebula. In SS433, the peculiarities of the
supercritical accretion onto black holes are most pronounced. Therefore, further multi-wavelength
studies of this unique microquasar are very promising.

\end{abstract}


\everymath{\displaystyle}

\section{Introduction}
\label{s:intro}

In astrophysics microquasars are usually called X-ray binary systems with accreting relativistic objects (neutron stars (NS) or black holes (BH)) and collimated ejections of matter moving with relativistic velocities (jets).
To date, several dozen microquasars have been discovered in the Galaxy. Ultraluminous X-ray sources (ULX) with luminosities up to $10^{42}$\,erg/s, which in many cases are similar in nature to microquasars, have been observed in other galaxies.

The value of microquasars for science is that in them in miniature the same processes occur as in the nuclei of active galaxies and quasars, which often demonstrate relativistic jets.
Both in quasars and microquasars,  accretion of matter onto a relativistic object (supermassive BH) occurs, only the source of matter in quasars is not a companion star in a binary system, but the stellar and gas environment of a supermassive BH in the galactic nucleus. 
Because of the large masses of supermassive BHs in quasars ($\sim10^5-10^{10}\,{\rm M}_\odot$), non-stationary processes in their vicinity occur on large timescales, which makes it difficult to elucidate their nature. At the same time, in microquasars all processes occur on relatively short timescales, which makes them very convenient objects for investigation.

The unique object SS433 is the first  microquasar discovered. In the Catalogue of stars with strong H$\alpha$ hydrogen emission lines compiled by Stephenson and Sanduleak \cite{1977ApJS...33..459S}, this object is listed under number 433. Hence its the name, SS433.
It is a strongly reddened due to interstellar absorption northern sky object of $\sim14$ stellar magnitude located in the centre of the W50 nebula, a relatively young ($10^4-10^5$ years old) peculiar supernova remnant (plerion, i.e. a supernova remnant that has an amorphous and elongated structure rather than a shell).

\begin{figure}
	\includegraphics[width=\columnwidth]{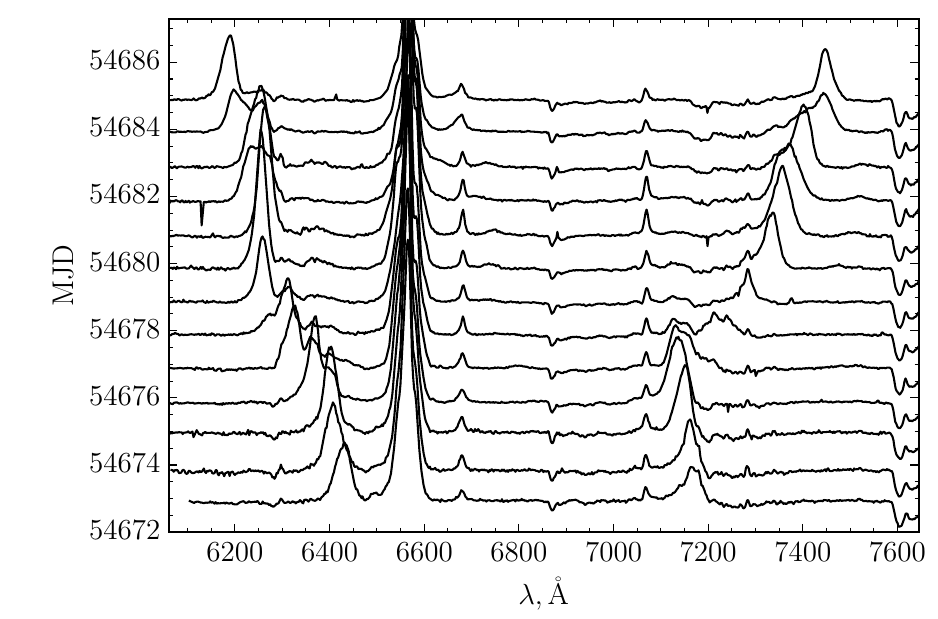}
    \caption{
    Sequence of SS433 spectra showing the stationary and moving hydrogen emission lines H$\alpha$ (6563\AA). The spectra are marked by modified Julian days (MJD, counted from midnight).}
    \label{f:spectra}
\end{figure}

In 1978, the journal Nature published a paper by British scientists D. Clark and P. Murdin \cite{1978Natur.276...44C}, in which they published optical spectra of a number of compact objects located in the centers of supernova remnants. SS433 attracted particular interest, in the spectrum of which, in addition to the standard emission lines of hydrogen and helium, there are many strong emission lines that can not be identified with any of the known chemical elements.
Further spectroscopic studies \cite{1979ApJ...230L..41M,1979ApJ...233L..63M,1980A&A....85...14M} have shown that there are two systems of emission lines of hydrogen and neutral helium in the spectrum of SS433: stationary, fixed emission lines with standard wavelengths, and moving emission lines that shift periodically across the spectrum with a period of $\sim$ 162 days by an enormous value, of the order of 1000 angstroms, which on the velocity scale corresponds to several tens of thousands of km/s, see Fig. \ref{f:spectra}. Nothing like this has ever been observed by astrophysicists before. The nature of SS433 seemed mysterious, so the object SS433 was dubbed ''The Enigma of the Century'' \cite{1981S&T....62..100S}.
In the first press releases there were even mentions that perhaps in this case we are observing signals from an extraterrestrial civilization beaming a super-powered tunable laser at us. Fortunately this hypothesis did not manage to get into scientific publications, because in 1979 it became clear\cite{1979A&A....76L...3M,1979MNRAS.187P..13F} that the moving emission lines in the spectrum of SS433 are produced in relativistic ($v\simeq 80\,000$ km/s) collimated (opening angle $\sim1^\circ$) jets that outflow from a central source and precess with a period of about 162 days. The opening angle of the precession cone is $\sim20^\circ,$ and the precession axis is tilted with respect to the line of sight by an angle of $\sim79^\circ.$
The moving emission lines are formed in the outer parts of the jets. The displacements of these lines along the SS433 spectrum are caused by the longitudinal Doppler effect, while the constant redshift of the center of symmetry of motion of the moving lines by $\sim 11\,000$ km/s is due to the transverse Doppler effect and reflects the relativistic time dilation in the moving matter of the jets. Although the nature of the central source remained unclear, it was speculated that all these phenomena occur in a binary system \cite{1979SvAL....5..344S}.

The model of a binary system was confirmed in 1980 in the work of Canadian scientists Crampton, Cowley, and Hutchings \cite{1980ApJ...235L.131C}, who measured the Doppler shifts of the narrow components of stationary hydrogen emissions and found a period of $\sim 13.1$ days. The authors, assuming that the radial velocity curve they measured reflects the orbital motion of the components, concluded from their data analysis that SS433 is a low-mass X-ray binary system consisting of a low-mass optical star ($\sim 1{\rm M}_{\odot}$) and a neutron star. As was shown later in our studies, this conclusion turned out to be incorrect.

In 1980, we performed photometric observations of SS433 \cite{1981MNRAS.194..761C} and found that this object is an eclipsing variable binary system. The discovery of optical eclipses in SS433 \cite{1981MNRAS.194..761C} allowed us to establish that the radial velocity curve measured by Crampton, Cowley, and Hutchings \cite{1980ApJ...235L.131C}, reflects not the orbital motion of binary components, but describes the motion of gas in the stream flowing from the inner Lagrangian point ${\rm L}_1$ onto the relativistic object, therefore it is impossible to estimate the masses of components from this radial velocity curve. The analysis of optical eclipses in SS433, as well as the study of the system's out-of-eclipse brightness as a function of the precessional 162-day period phase \cite{1981MNRAS.194..761C} together with information on the interstellar absorption for SS433 and its luminosity \cite{1980MNRAS.193..135M} allowed us to construct a self-consistent, now generally accepted model of SS433.

According to \cite{1981MNRAS.194..761C}, SS433 is a massive eclipsing X-ray binary system at an advanced stage of evolution, where an optical donor star overflows its Roche lobe and transfer matter onto a relativistic object at a very strong rate of $\sim10^{-4}\,{\rm M}_{\odot}/$year. Because of the enormous accretion rate onto the relativistic object, an accretion disk formed around it is opaque to the generated X-ray radiation. Therefore, accretion proceeds in the supercritical regime predicted by N.I. Shakura and R.A. Sunyaev \cite{1973A&A....24..337S}. The supercritical accretion disk is an optically bright object. Relativistic jets are launched perpendicular to the central parts of the accretion disk and track the 162-day precession of the disk plane, which is tilted to the orbital plane by an angle of $\sim$ 20$^\circ$. The moving emissions are formed in the outer parts of the jets.

The massive X-ray binary system model for SS433 proposed in \cite{1981MNRAS.194..761C} was confirmed in  \cite{1981ApJ...251..604C}, in which the authors constructed a radial velocity curve from the HeII 4686\AA\ stationary emission line and measured the relativistic mass function (a lower limit for the mass of an optical star), which was found to be close to $10\,{\rm M}_{\odot}$. The question arose about the causes of the precession of the accretion disk. As early as in 1973, Katz \cite{1973NPhS..246...87K} proposed a model of a slaved precessing disk to explain the 35-day X-ray cycle of the Her X-1 binary system (for a modern precession model for Her X-1, see \cite{1999A&A...348..917S}).

In 1974, Roberts \cite{1974ApJ...187..575R} proposed a slaved (floating) precession model for the accretion disk in the Her X-1 system. Even earlier, in 1973, the idea of a slaved disk tracking the precession of the rotation axis of an optical matter donor star was proposed by Shakura \cite{1973SvA....16..756S}.
In 1980, the slaved disk model was applied to analyze observations of SS433 \cite{1980A&A....81L...7V}. In this model, the rotational axis of the optical star precesses under the gravitational attraction of the relativistic object, and the accretion disk, which does not lie in the orbital plane, tracks this precession. The non-perpendicularity of the optical star's rotational axis to the orbital plane may be related to the asymmetric supernova explosion, which turned the orbital plane of the binary system relative to the optical star's rotation axis \cite{1974ApJ...187..575R,1981SvAL....7..401C}. To explain the high stability of the matter velocity in SS433 jets ($80\,000$ km/s), a model of radiation acceleration of matter was proposed (the so-called line-locking effect,  \cite{1982IAUS...97..209S,Shapiro86}).
Thus, by the early 1980s, the main features of SS433 as a massive eclipsing X-ray double system in the supercritical accretion regime had been elucidated. However, many questions remained unresolved. Here are the main ones.

\begin{enumerate}
    \item 
The nature of SS433 precession variability. Which model is applicable to SS433: the tidally precessing disk model\cite{1973NPhS..246...87K} or the slaved disk model \cite{1973SvA....16..756S}?

 \item 
 The nature of the accreting relativistic object: neutron star or black hole?
 
\item 
Why, contrary to theoretical predictions, has a common envelope not formed in SS433  experiencing secondary mass exchange on a thermal time scale and the system evolves as a semi-detached binary?

\item
Is the formation of relativistic jets related to the supercritical accretion regime, or they are formed by other mechanisms?

\item 
Structure and features of the supercritical accretion disk around a relativistic object in SS433, including a powerful stellar wind from it.

\end{enumerate}


\section{Modern status}

Object SS433 is the first example of a microquasar in our Galaxy. It has been studied for more than 45 years in almost all ranges of the electromagnetic spectrum: optical, infrared, radio, X-ray, and gamma-ray (see reviews \cite{2004ASPRv..12....1F,2020NewAR..8901542C} and references therein). It is a massive X-ray binary system at an advanced evolutionary stage with a precessing supercritical accretion disk and relativistic jets. As mentioned above, the object is located at the center of the relatively young ($10^4-10^5$ years old) supernova remnant W50  at a distance of $\sim 5$ kpc\cite{1981MNRAS.194..761C,1984ARA&A..22..507M,2004ASPRv..12....1F}. 
    
The binary system SS433 exhibits three types of regular spectral and photometric variability: precessional ($P_{\rm prec} \simeq162.3^d$), orbital ($P_{\rm orb}\simeq 13.1^d$), and nutational ($P_{\rm nut}\simeq6.29^d$) ones. The precessional variability is related to the change of orientation of the optically bright supercritical accretion disk and relativistic jets relative to the observer. The orbital variability is due to eclipses of the components: the disk by the companion star and the star by the disk. The nutational variability is related to the wobbling of the disk plane and jets under the influence of tidal forces acting on the tilted precessing disk from the orbiting optical star. The orbital, precessional, and nutational periods are related by a formula that takes into account the addition of frequencies of the precessional and nutational variability $P_{\rm nut}^{-1}=2P_{\rm orb}^{-1}+P_{\rm prec}^{-1}$.
The stability of the precessional period over 45 years (except for some phase jumps) favors the slaved accretion disk model 
\cite{1974ApJ...187..575R,2022ARep...66..451C}, because due to the huge moment of inertia of the optical star, the period of forced precession of its rotational axis should be very stable.
The model of precessing jets proposed in \cite{1979A&A....76L...3M,1979MNRAS.187P..13F} was brilliantly confirmed by direct observations of jets in the radio band with high angular resolution \cite{1981Natur.290..100H,1988A&A...189..124F}.
    
Moving emission lines of various chemical elements were also detected in X-rays \cite{1994PASJ...46L.147K}. Radio interferometric observations with a milliarcsecond angular resolution \cite{2001ApJ...562L..79B} detected an equatorial outflow of gas from SS433, perpendicular to the directions of the jets and propagating from SS433 to a distance of several hundred astronomical units. This suggests that the optical star in SS433 overflows its Roche lobe and the mass transfer occurs not only through the inner Lagrangian point L$_1$ but also through the outer Lagrangian point L$_2$. Additional arguments pointing at such an outflow and the presence of a corresponding circumstellar shell around SS433 rotating with a velocity of about 200 km/s are given in papers \cite{1993MNRAS.261..241F,1988AJ.....96..242F,2011A&A...531A.107B}.

The W50 nebula, centered on SS433, has been studied in the optical, radio, X-ray, and gamma-ray bands (see, for example, \cite{1998AJ....116.1842D,2007A&A...474..903B}). The interaction of relativistic jets with the matter of the nebula determines its elongated shape and the presence of gas knots (filaments) in the collision sites of jets with the interstellar medium.
	
An important and difficult problem in the case of SS433 is the elucidation of the nature of the relativistic object: a neutron star or a black hole? The supercritical optically bright accretion disk in SS433 makes it difficult to detect absorption lines in the spectrum of the optical star, so for a long time the mass function of the optical star was unknown. It was not until more than twenty years after the discovery of SS433 that the absorption lines in the spectrum of the optical star were recorded, their Doppler shifts were measured, and the spectral type of the star was determined, which turned out to be an A7I  supergiant \cite{2002ApJ...578L..67G,2004ApJ...615..422H,2008ApJ...676L..37H}. The semi-amplitude of the optical star's radial velocity curve is $K_{\rm v}=58.2\pm3.1$ km/s,
the corresponding mass function of the optical star $f_{\rm v} (M)=(M_{\rm x}^3\sin^3i)/(M_{\rm x}+M_{\rm v} )^2 = 0.268\,{\rm M}_\odot$.
Similar results were obtained in \cite{2010ApJ...709.1374K}. In the recent paper \cite{2020A&A...640A..96P}, along with the construction of the radial velocity curve of the optical star in the SS433 system, the axial rotation velocity of the optical star $v_\mathrm{rot}=140\pm20$ km/s was measured. This allowed the authors to estimate the component mass ratio $q=M_{\rm x}/M_{\rm v} = 0.37 \pm 0.04$ and, given the orbital inclination of the system known from the observations of mobile emissions $i = 79^\circ$, to determine the component masses from the observed mass function of the optical star: $M_{\rm x} = 4.2\pm 0.4\,{\rm M}_\odot$, $M_{\rm v} = 11.3\pm 0.6\,{\rm M}_\odot$, where $M_{\rm x}$ and $M_{\rm v}$ are masses of the relativistic object and optical star, respectively.
    
It can be assumed that the mass estimates of the relativistic object in SS433 based on the spectral data on the absorption lines and the optical star's radial velocity curve cannot be considered as reliable because the radiation from the donor star passes through the moving gas inside the binary system, as well as through the circumbinary envelope rotating at a velocity of about 200 km/sec. At the enormous mass-loss rate of $\dot M \sim 10^{-4}\,{\rm M}_\odot$/year, the matter density in these structures is very large. Therefore, selective absorption of light from the optical star in dense moving gas structures can significantly distort the orbital radial velocity curve constructed from absorption lines of the donor star (see, e.g., \cite{2008ApJ...678L..47B,2013A&A...556A.149B,2018A&A...619L...4B}).
    
The situation with stationary emission lines in the spectrum of SS433 is also uncertain. For example, the mass function
of the relativistic object $f_{\rm x}(M)=(M_{\rm v}^3\sin^3 i)/(M_{\rm x}+M_{\rm v} )^2$, determined from the Doppler shifts of the stationary HeII 4686\AA emission, depends on the phase of the precessional period and varies from $\sim 10\,{\rm M}_\odot$ \cite{1981ApJ...251..604C} to $\sim 2\,{\rm M}_\odot$ \cite{1991Natur.353..329D}. This may be due to the complex structure of the high-speed wind from the precessing supercritical accretion disk. The shape of this disk may be twisted and asymmetric, and the wind region in which the stationary HeII 4686\AA emission is formed may not accurately reflect the orbital motion of the relativistic object. At wind speeds of the order of $1-2$ thousand km/s and the observed semi-amplitude of the radial velocity curve constructed from the HeII 4686\AA emission which, according to various studies, ranges from 112 km/s \cite{1991Natur.353..329D} to $195$ km/s \cite{1981ApJ...251..604C}, a small ($\sim 10\%$) wind asymmetry can introduce into the true radial velocity curve errors comparable to the value of the true semi-amplitude of the radial velocity curve. Therefore, the mass estimates of the relativistic object inferred from the stationary emission lines in SS433 also cannot be considered reliable.
    
Due to the described uncertainty, great hopes were placed on the results of studies of X-ray eclipses in SS433. Since the orbital inclination of SS433 $i = 79^\circ$ is reliably known from the analysis of periodic displacements of moving emissions, and the optical donor star fills or overflows its Roche lobe (whose size depends on the component mass ratio $q=M_{\rm x}/M_{\rm v}$ ), the analysis of duration of the X-ray eclipse (when the optical star eclipses the central X-ray source or the bases of relativistic jets) enables the estimation the binary mass ratio and then,  from the mass function of the optical star or relativistic object, the mass of the relativistic object.
The first such estimates were made in papers \cite{1989PASJ...41..491K,1989A&A...218L..13B,1998IAUS..188..358K}. From the analysis of X-ray eclipses in SS433 in the $2-10$ keV range, using the model of thin relativistic X-ray emitting jets, the authors obtained a relatively small mass ratio $q\simeq 0.15$. It was assumed that the optical donor star exactly fills its Roche lobe.
    
However, there is observational evidence that the optical star in the system not only fills, but even overflows its Roche lobe and transfers mass not only through the inner Lagrangian point ${\rm L}_1$, but also through the outer Lagrangian point ${\rm L}_2$. For example, evidence that the optical star outflows through the ${\rm L}_2$ point has been obtained from analyzing spectral data  \cite{1993MNRAS.261..241F}.
This means that the star fills the outer Roche lobe, whose size is $15-20$\% larger than the size of the inner critical Roche lobe. As noted above, radio astronomical methods have discovered an equatorial outflow of matter from SS433, perpendicular to the direction of radio jets, which extends from SS433 to a distance of hundreds of astronomical units \cite{2001ApJ...562L..79B}.
This also suggests the matter outflows through the external Lagrangian point ${\rm L}_2$ and the optical star fills its external Roche lobe. This conclusion is also supported by optical and IR spectroscopic observations of SS433, which revealed the presence of double-humped stationary hydrogen lines in its spectrum. These double-humped emission lines indicate the presence of a circumstellar envelope in SS433, which was most likely formed as a result of the matter outflow through the outer Lagrangian point ${\rm L}_2$.
	
Recently, theoretical studies \cite{2015MNRAS.449.4415P,2017MNRAS.465.2092P} showed that due to a limited width for the outflowing  matter in the vicinity of the inner Lagrangian point ${\rm L}_1$ from a star with a radiative envelope during the thermal scale mass transfer process in a binary system, the star can significantly overflow its inner Roche lobe for a long time, and the star's matter can outflow through the outer Lagrangian point ${\rm L}_2.$
    
Thus, since the radius of the eclipsing star in SS433 is much larger than the size of the inner Roche critical lobe, the estimate of the mass ratio $q\simeq0.15$ obtained in 
\cite{1989PASJ...41..491K,1989A&A...218L..13B,1998IAUS..188..358K} from the analysis of the X-ray eclipse duration should be taken as a lower limit: $q > 0.15$.
    
In paper \cite{2020NewAR..8901542C}, we presents the results of interpretation of the orbital eclipsing and precessional light curves  of SS433 in the hard X-ray spectrum observed by the INTEGRAL space observatory.
It is shown that since the X-ray spectral hardness does not change with orbital and precessional phases of the system (while the observed X-ray flux changes by a factor of five), the hard X-ray emission ($kT=18-60$ keV), in contrast to the soft X-ray emission ($kT=2-10$ keV), should be formed not in narrow collimated jets, but in an extended quasi-isothermal corona above the supercritical accretion disk.
According to calculations \cite{2009MNRAS.394.1674K}, the corona temperature is $kT\sim 0.2$ keV, the mass loss rate in relativistic jets is $\dot M_j = 3\times 10^{-7}\,{\rm M}_\odot$/year, and the mass loss rate in the disk wind is $\dot M\sim 10^{-4}\,{\rm M}_\odot$/year.

In paper \cite{2020NewAR..8901542C}, in the framework of the model of SS433 as an X-ray binary system with an optical star filling the outer Roche lobe, an estimate of the mass ratio $q>0.4\div 0.8$ was obtained from a joint analysis of the eclipsing and precessional hard X-ray variability of the system. In this analysis, the value $q>0.4$ corresponds to a model in which the third not eclipsed light from the X-ray radiation scattered in the extended wind region from the supercritical accretion disk is taken into account, and in the case of $q>0.8$ the contribution of the third not eclipsed light is assumed to be zero.

Summarizing, we can conclude that based on  the radial velocity curves of the components of SS433 it is not possible to obtain reliable values of the component masses, and the analysis of X-ray eclipses in this system allows us to obtain reasonable results only under a special model assumption about that the donor star fills its external Roche lobe.

In this review, we present estimates of the mass ratio and component masses of SS433 from independent observational data. A 30-year spectral and photometric monitoring of SS433 has been organized at Stermberg Astronomical Institute of Moscow State University (SAI MSU). The analysis of these data together with published data enabled us to trace the evolution of SS433 over 45 years and to obtain independent estimates of parameters of this microquasar.

\section{Spectral monitoring}

\begin{figure}
	\includegraphics[width=\columnwidth]{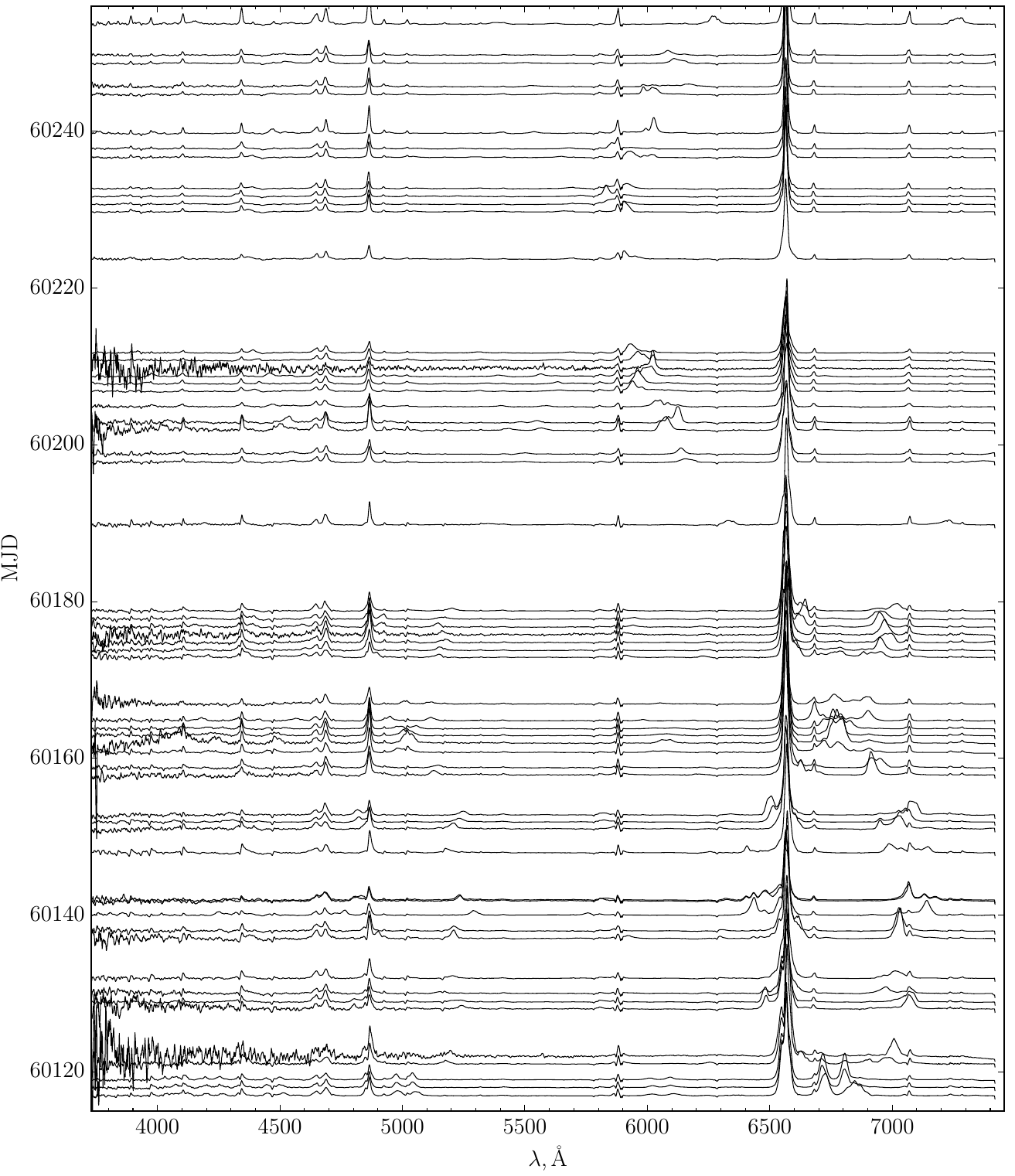}
    \caption{
 Stationary and moving emissions in the spectrum of SS433. Sequence of SS433 spectra obtained in summer-autumn 2023 at the Caucasus Mountrain Observatory of SAI MSU. On the ordinate axis -- time in modified Julian days.
    }
    \label{f:tds}
\end{figure}

\begin{figure}
	\includegraphics[width=\columnwidth]{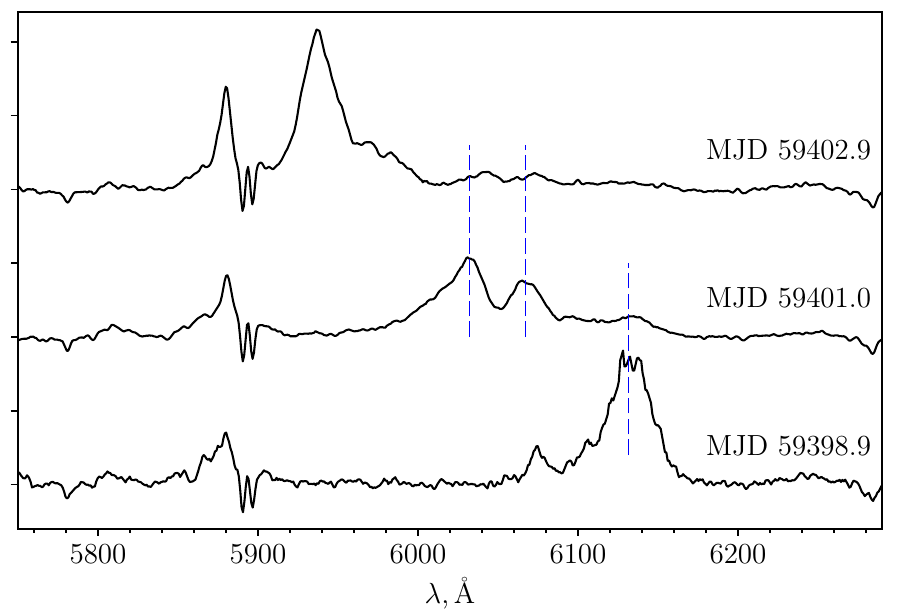}
    \caption{
Structure of moving H$\alpha$ emissions formed in relativistic jets. Three spectra of SS433 obtained with an interval of two days. One can see the effect of gradual fading of the moving line and the multicomponent structure of the moving lines, partly explained by the fading effect. The moving line component and its residual afterglow are connected by blue dashed lines.
    }
    \label{f:lines}
\end{figure}


Since 1994, spectral monitoring of SS433 has been carried out at the SAI MSU telescopes to check the stability of the object's parameters and to detect evolutionary effects in this massive X-ray binary system at an advanced evolutionary stage. The observations are carried out at the Crimean Astronomical Station (CAS) of SAI MSU (1.25-meter ZTE telescope) and at the Caucasus Mountain Observatory (CMO) of SAI MSU (2.5-meter telescope). The Crimean observations have been carried out on the A-spectrograph designed by  V.F. Esipov  with a spectral resolution of $R = \delta\lambda/\lambda \approx 1000$, and CMO observations have been carried out on the two-channel Transien Double-beam Spectrograph (TDS, $3600 \div 7400$\AA, spectral resolution $\sim 2000$)
\cite{2020AstL...46..836P}). The  results of monitoring of SS433 from 1994 to 2021 and description of instrumentation and data processing methods  are published in 
\cite{2008ARep...52..487D,2018ARep...62..747C,2022ARep...66..451C}. Here we complement these observations with new observations carried out in $2021-2024$ and present results of their joint analysis.    

\subsection{Kinematical model of SS433}
\begin{figure}
	\includegraphics[width=\columnwidth]{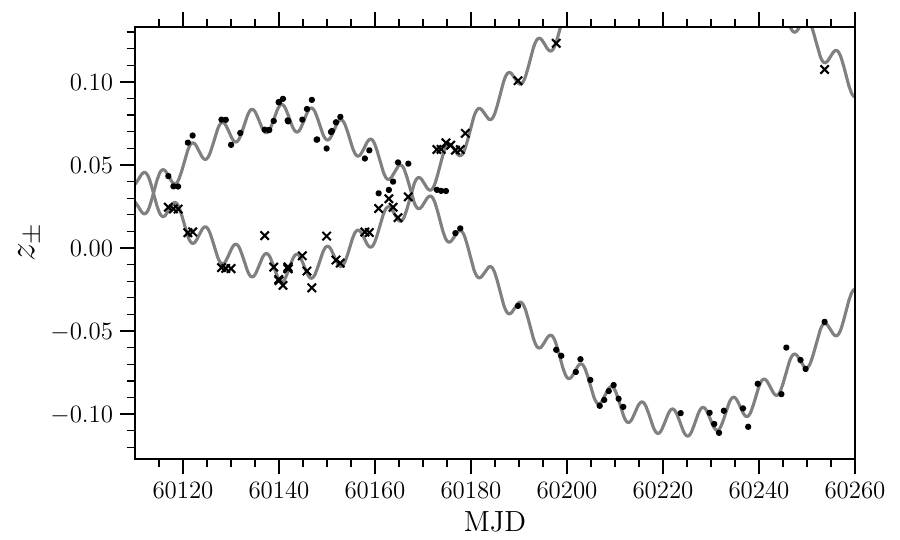}
    \caption{
    Precession-nutation displacement of moving lines. The time interval corresponds to Fig. \ref{f:tds}. Gray curve corresponds to the kinematic model with parameters found from the time interval shown. The Doppler shifts of the blue and red H$\alpha$ emissions as a function of time are plotted on the ordinate axis.
    }
    \label{f:precnut}
\end{figure}
Fig. \ref{f:tds} shows a sequence of SS433 spectra obtained on close observational nights in the summer -- fall of 2023, which demonstrates stationary H$\alpha$ emission as well as moving redshifted  (H$\alpha^+$) and blueshifted (H$\alpha^-$) emissions .
The moving emissions have a complex time-dependent multicomponent structure. This reflects the fact that matter in the jets is concentrated in separate clumps whose luminosity in the H$\alpha$ line frequencies fades as the clumps move away from the center of the supercritical accretion disk (see Fig. \ref{f:lines}). The characteristic time of $\delta t$ in moving emission lines is not constant and ranges from a fraction of a day to several days, which imposes constraints on the matter density in jets $n_{\rm e}\sim 13/\delta t\sim 10^8$ cm$^{-3}$.
Fig. \ref{f:precnut} shows a fragment of the radial velocity curve measured from the moving H$\alpha^+$ and H$\alpha^-$ emissions for the same time interval as in Fig. \ref{f:tds}. These curves clearly display the nutational variability of the radial velocities with period  $P_\mathrm{nut}\simeq 6^d.29$, which is superimposed on the precessional variability with period $P_\mathrm{prec} = 160$ days. The center of symmetry of the motion of moving emissions is redshifted relative to the stationary H$\alpha$ emission by $\sim 200${\AA} (in the velocity scale $\sim 11\,000$ km/s), which, as already noted, is due to the transverse Doppler effect.

The Doppler shifts of the moving lines are usually described by a simple geometric model that includes precession and nutation changes in the direction of the jet perpendicular to the accretion disk plane and assumes that the jet and counter-jet move along the same straight line with the same velocity \cite{1979Natur.279..701A,1984ARA&A..22..507M}:

\begin{equation}
\begin{split}
  z_{\pm}+1         &=z_{\rm prec\pm}+z_{\rm nut\pm }                \\
  z_{\rm prec\pm}      &=\gamma [  1 \pm \beta \sin i \sin \theta \cos \varphi_{\rm prec}(t) \pm \beta \cos i \cos \theta ]        \\    
  z_{\rm nut\pm}    &= \pm Z_{\rm nut}\cos\varphi_{\rm nut}(t)                                        \\
  \varphi_{\rm prec}(t)           &=  2\pi(t-t_0)/P_{\rm prec},    \\
  \varphi_{\rm nut}(t) &= 2\pi(t-t_{\rm nut})/P_{\rm nut}\,.   
\end{split}
\end{equation}
%
Here $z_{\pm}=\lambda_{\pm}/\lambda_0-1$ is the observed line shift, $z_{\rm prec\pm}$ is the component that accounts for the transverse Doppler effect and precession, $z_{\rm nut\pm}$ is the correction for nutation, $\beta=v/c$ is the dimensionless jet velocity, $i$ is the inclination of the precessional axis to the line of sight, $\theta$ is the opening angle of the precession cone, $\gamma = 1/\sqrt{1-\beta^2}$ is the  Lorentz factor, $P_{\rm prec}$ is the precessional period, $t_0$ is the initial epoch when the positions of both line components are maximally displaced (ignoring nutation), $Z_{\rm nut}$ is the amplitude of change of $z_{\rm nut\pm}$, $P_{\rm nut}$ is the nutational period, $t_{\rm nut}$ is the initial epoch to describe nutation.

The unknown model parameters $\beta, \theta, i, P_{\rm prec},  t_0, Z_{\rm nut},  P_{\rm nut},  t_{\rm nut}$ and their errors are determined by least-square method using the Levenberg-Marquardt  algorithm \cite{1992nrfa.book.....P}.
Spectral observational data (including both our new and previously published data) obtained over 45 years from 1978 to 2023 were used. The total number of measured positions of the moving lines is 2239 and 1926 for the blue and red components, respectively. The methods and results of solving our inverse problem are described in 
\cite{2008ARep...52..487D,2018ARep...62..747C,2022ARep...66..451C}
As the SS433 precessional variability shows jumps in the initial phase (see below), we determined the kinematic model parameters using relatively short time intervals of $2-3$ years.
In the case where both components (red and blue lines) are measured together, one can directly determine the $\beta$ parameter for each observation by the transverse Doppler effect (assuming symmetrically outflowing jets):
\begin{equation}
    \beta=\sqrt{1-\left(1+\frac{z_-+z_+}{2}\right)^{-2}}
    \label{e:beta}
\end{equation}
Such calculations show that $\beta(t)$ is not a constant and can vary with time in the interval $0.21 \div 0.30$. However, taking into account this variability when determining the model parameters does not increase the accuracy because it decreases the number of observations with simultaneous measurement of both components of the moving lines, which is not always possible (e.g., due to superposition of the moving line on the stationary line H$\alpha$ or absorption bands of the terrestrial atmosphere, limited spectral range of the TDS spectrograph, etc.). For this reason, the parameters of the kinematic model were determined by keeping $\beta$ constant over the time interval considered.

\subsection{Parameters of the kinematic model}

\begin{figure}
	\includegraphics[width=0.75\columnwidth]{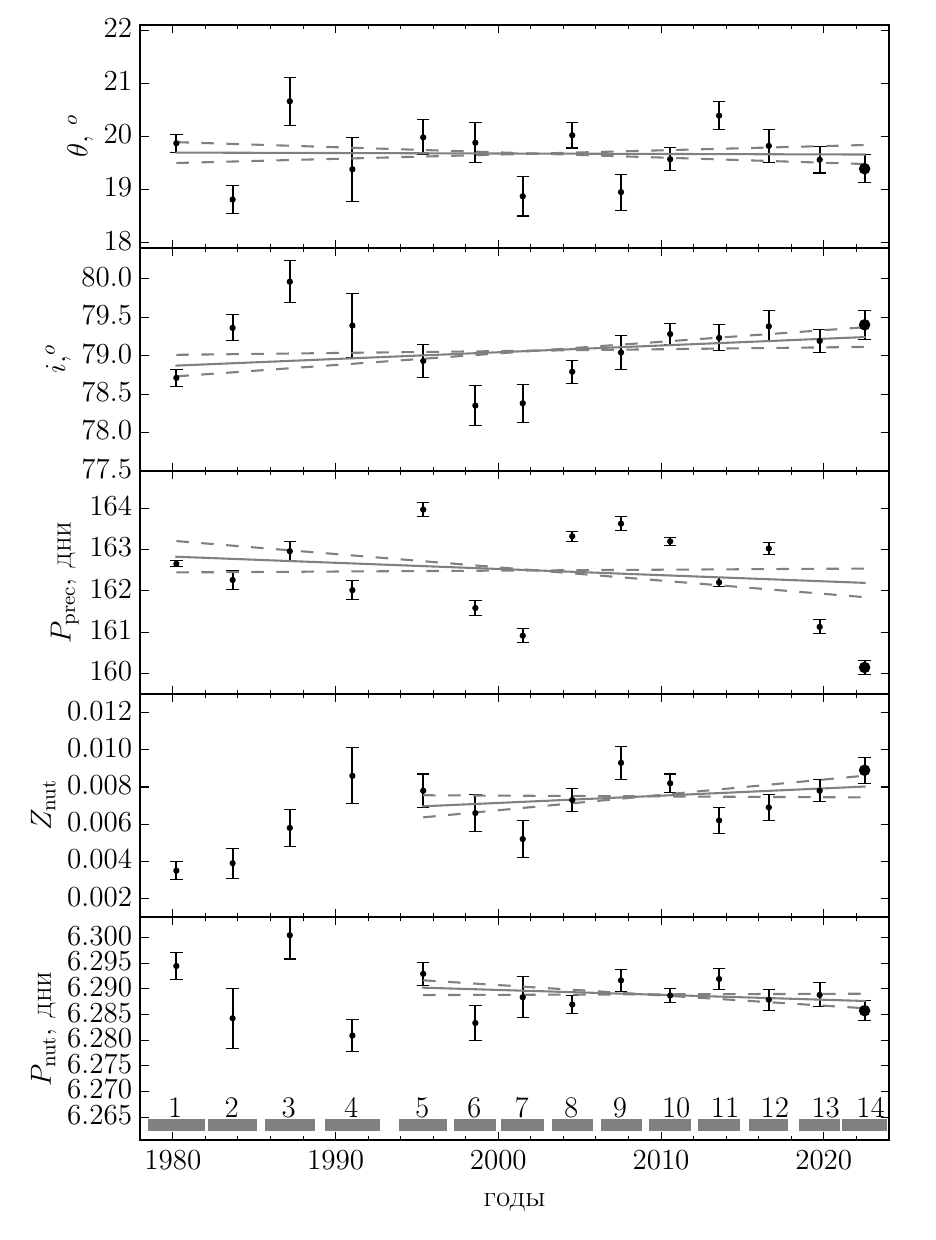}
    \caption{
    Parameters of the kinematic model of SS 433 (see Tables \ref{t:precpar} and \ref{t:nutpar}). The bold dots on the right (the 14th interval) are derived from new data.
    }
    \label{f:param}
\end{figure}

\begin{table*}
\caption{
Precessional parameters of the model. Average values for 45 years are given at the bottom.
}
 \label{t:precpar}
\footnotesize
\begin{tabular}{c | c | c | c | c | c | c  }
\hline
N & $\rm JD_{start}$ -- $\rm JD_{end},$ & $\beta$ & $\theta,$ & $i,$ & $P_{\rm prec},$     & $t_0,$              \\
  &           days                    &         &   $^\circ$    & $^\circ$ & days              & days               \\
\hline  
1  & 43688 --  44946 & 0.2602\,(0.0013) & 19.87\,(0.17) &  78.71\,(0.11) & 162.65\,(0.07)  &  44320.49\,(0.2)     \\
2  & 45048 --  46114 & 0.2621\,(0.0021) & 18.81\,(0.27) &  79.36\,(0.17) & 162.26\,(0.22)  &  45295.15\,(0.5)     \\
3  & 46320 --  47408 & 0.2578\,(0.0028) & 20.66\,(0.45) &  79.96\,(0.27) & 162.96\,(0.23)  &  46919.48\,(0.7)     \\
4  & 47668 --  48870 & 0.2594\,(0.0046) & 19.38\,(0.60) &  79.39\,(0.42) & 162.02\,(0.23)  &  48376.11\,(0.8)     \\
5  & 49336 --  50381 & 0.2568\,(0.0026) & 19.98\,(0.33) &  78.93\,(0.21) & 163.96\,(0.17)  &  50000.08\,(0.5)     \\
6  & 50570 --  51488 & 0.2586\,(0.0028) & 19.88\,(0.38) &  78.35\,(0.26) & 161.58\,(0.18)  &  51134.20\,(0.4)     \\
7  & 51626 --  52560 & 0.2591\,(0.0027) & 18.87\,(0.37) &  78.38\,(0.25) & 160.91\,(0.18)  &  52104.14\,(0.5)     \\
8  & 52754 --  53654 & 0.2558\,(0.0018) & 20.02\,(0.24) &  78.79\,(0.15) & 163.32\,(0.13)  &  53244.92\,(0.4)     \\
9  & 53850 --  54748 & 0.2620\,(0.0027) & 18.95\,(0.34) &  79.04\,(0.22) & 163.62\,(0.18)  &  54381.39\,(0.5)     \\
10 & 54944 --  55860 & 0.2601\,(0.0016) & 19.57\,(0.22) &  79.28\,(0.14) & 163.20\,(0.11)  &  55517.31\,(0.3)     \\
11 & 56043 --  56958 & 0.2575\,(0.0020) & 20.39\,(0.27) &  79.23\,(0.17) & 162.21\,(0.11)  &  56656.28\,(0.3)     \\
12 & 57186 --  58046 & 0.2579\,(0.0023) & 19.82\,(0.31) &  79.38\,(0.20) & 163.02\,(0.14)  &  57632.05\,(0.5)     \\
13 & 58314 --  59202 & 0.2600\,(0.0020) & 19.56\,(0.25) &  79.19\,(0.15) & 161.13\,(0.17)  &  58935.07\,(0.3)     \\
14 & 59279 --  60254 & 0.2578\,(0.0023) & 19.39\,(0.26) &  79.40\,(0.19) & 160.14\,(0.17)  &  59898.78\,(0.4)     \\
\hline                                                                                    
    &43688 --  60254 & 0.2591\,(0.0007) & 19.64\,(0.10) &  78.92\,(0.06) & 162.310 0.002   &  51298.908\,(0.13)   \\
\hline
\end{tabular}
\end{table*} 

\begin{table}
\caption{
Nutational parameters of the model. Average values for 45 years are given at the bottom.
}
 \label{t:nutpar}
\footnotesize
\begin{tabular}{c|c|c|c}
\hline
N  &  $Z_{\rm nut},$  & $P_{\rm nut}$   & $t_{\rm nut}$ \\ 
   &  $10^{-3}\times$  & days           &  days          \\ 
\hline
1  & 3.5\,(0.5)  & 6.2945\,(0.0026) & 44297.84\,(0.15) \\   
2  & 3.9\,(0.8)  & 6.2843\,(0.0058) & 45298.02\,(0.21) \\   
3  & 5.8\,(1.0)  & 6.3005\,(0.0046) & 46832.07\,(0.18) \\   
4  & 8.6\,(1.5)  & 6.2809\,(0.0031) & 48259.57\,(0.17) \\   
5  & 7.8\,(0.9)  & 6.2929\,(0.0023) & 49988.76\,(0.11) \\   
6  & 6.6\,(1.0)  & 6.2834\,(0.0034) & 51032.27\,(0.15) \\   
7  & 5.2\,(1.0)  & 6.2884\,(0.0040) & 52082.07\,(0.20) \\   
8  & 7.3\,(0.6)  & 6.2870\,(0.0018) & 53207.55\,(0.08) \\   
9  & 9.3\,(0.9)  & 6.2917\,(0.0022) & 54321.00\,(0.10) \\   
10 & 8.2\,(0.5)  & 6.2887\,(0.0013) & 55534.53\,(0.06) \\   
11 & 6.2\,(0.7)  & 6.2919\,(0.0021) & 56509.28\,(0.11) \\   
12 & 6.9\,(0.7)  & 6.2879\,(0.0021) & 57653.64\,(0.11) \\   
13 & 7.8\,(0.6)  & 6.2888\,(0.0024) & 58949.24\,(0.08) \\     
14 & 8.9\,(0.7)  & 6.2858\,(0.0020) & 59797.68\,(0.09) \\     
\hline
   & 6.3\,(0.3)  & 6.28802\,(0.00005) & 51126.64  0.05  \\
\hline
\end{tabular}
\end{table}

The parameters of the kinematic model are summarized in Tables \ref{t:precpar} and \ref{t:nutpar} and are shown in Fig. \ref{f:param}.  The last point in Fig. \ref{f:param} corresponds to the time interval from 2021 to 2023. It can be seen that all parameters of the kinematic model are stable on average over 45 years and do not show noticeable secular changes. This supports the model of a slaved accretion disk tracking the precession of the rotation axis of the optical star
\cite{1974ApJ...187..575R,1973SvA....16..756S,1980A&A....81L...7V}. 
On short time intervals of the order of several years, some deviations of the kinematic model parameters from the mean values are observed, but, with the exception of the $P_\mathrm{prec}$ parameter, these deviations do not exceed $2-3$ $\sigma$ and are most likely due to the complex and variable shape of the profiles of moving emissions (see Fig. \ref{f:lines}), which makes it difficult to precisely determine the positions of these emissions. In the case of the precession period $P_\mathrm{prec}$, significant deviations of the instantaneous values of $P_\mathrm{prec}$ from the mean value are observed. As shown in paper \cite{2022ARep...66..451C}, these deviations reflect phase jumps in the precessional variability.

\subsection{Jumps in the precessional phases}

\begin{figure}
	\includegraphics[width=\columnwidth]{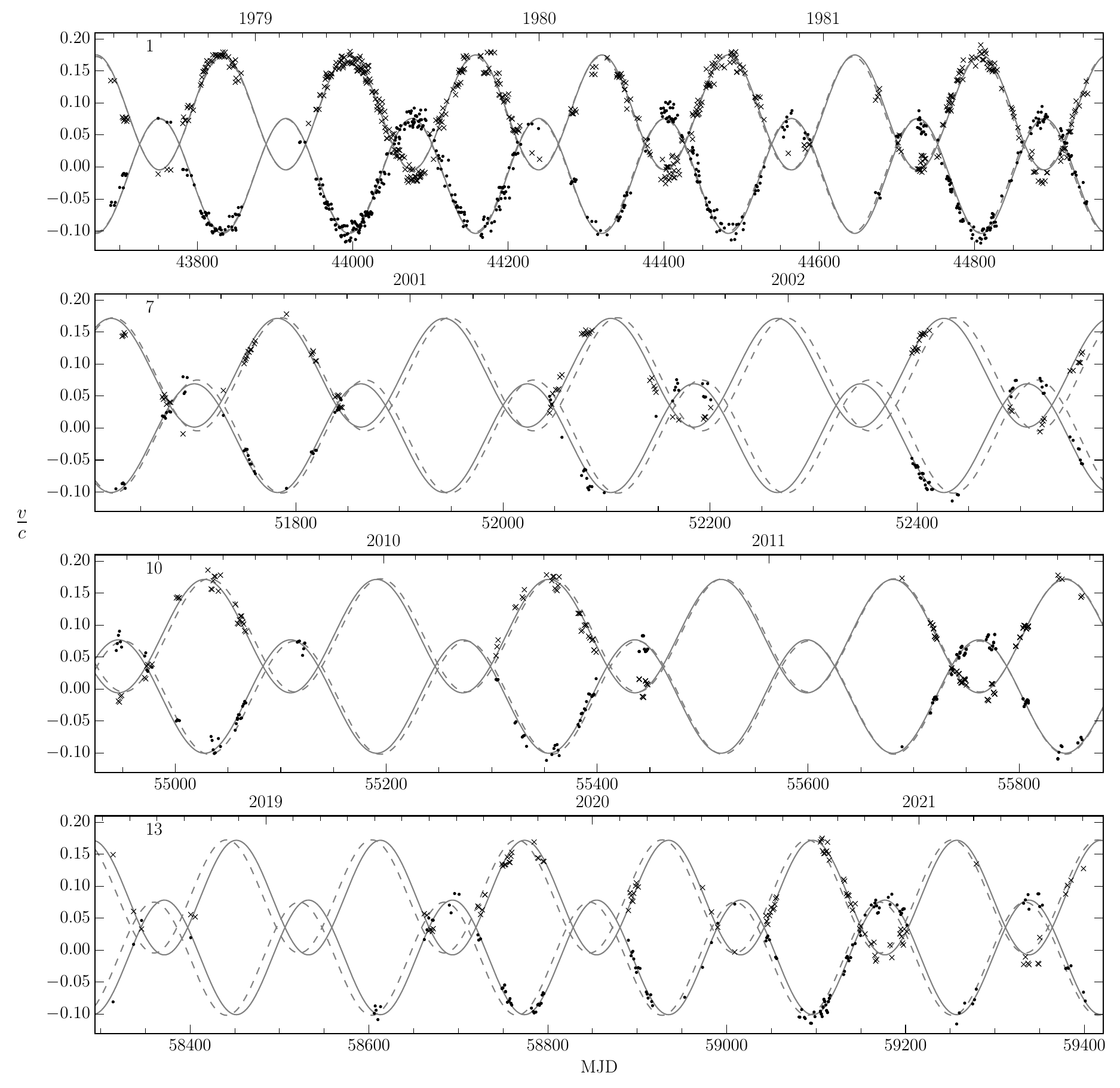}
    \caption{
    The observed radial velocities of the moving lines $v/c$ for individual three-year intervals are shown as dots (H$\alpha^-$) and crosses (H$\alpha^+$). The solid line corresponds to the kinematic model constructed using data from each interval separately. The dashed line corresponds to the kinematic model constructed from all available data (from \cite{2022ARep...66..451C}).
    }
    \label{f:vr_prec}
\end{figure}

\begin{figure}
	\includegraphics[width=\columnwidth]{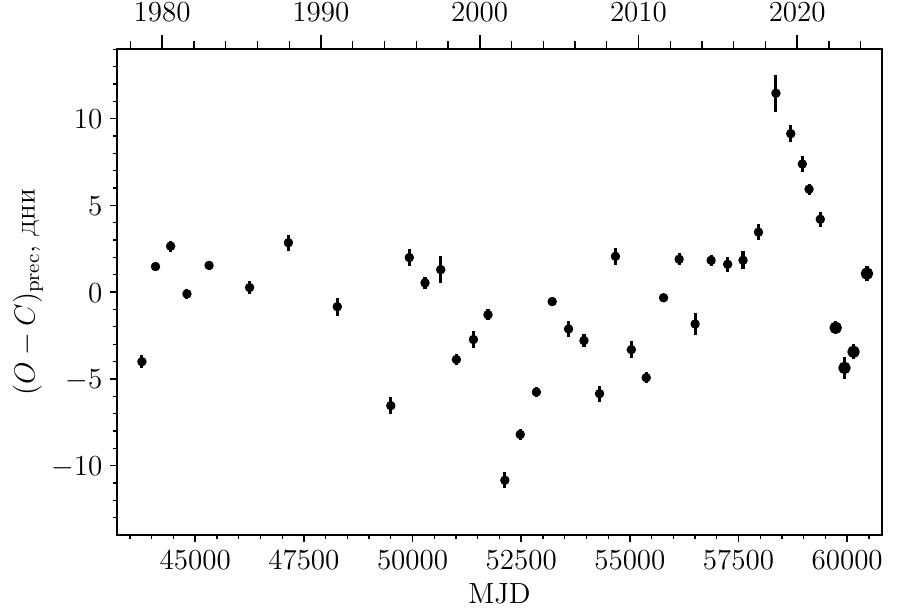}
    \caption{
    Observational season-averaged shifts in the phase of precessional variability relative to the mean kinematic model for all data, including new observations from the $2022-2044$ season (rightmost points, bolded).
    }
    \label{f:OCprec}
\end{figure}

As seen from Fig. \ref{f:param}, the observed value of the precessional period in the absence of secular variations sometimes deviates from the mean value of $162^d.3$ by an amount reaching $\sim \pm2^d$ ($\sim$1\% of $P_\mathrm{prec}$). These phase jumps can be seen in Fig. \ref{f:vr_prec}, where a comparison of the observed radial velocities of moving emissions with the average kinematic model constructed from  45-years observations is shown for several time intervals. This figure suggests that the phase deviations of the precessional variability occur on timescales comparable to the length of the intervals themselves. As each interval encompasses several precessional periods, we can find the mean deviation of the observed curve from the mean value for each observational season by fixing the parameters of the kinematic model at their average values. The time dependence of such deviations is shown in Fig. \ref{f:OCprec}, where there are two striking large episodes: lagging (around 2002) and advance (2019) in phase from the mean curve by $\sim 11$ days. Both deviations occurred on timescales of less than 1 year and disappeared within $2-3$ years.

In the slaved disk model, the propagation of relativistic jets tracks the precessing rotation axis of the donor star. However, precessional changes in the direction of this axis are reflected on the direction of jets not instantaneously, but after some time, which is necessary for the matter with a new angular momentum vector to reach the formation site of relativistic jets, i.e., after about viscous time of the accretion disk. In such a model, the observed phase fluctuations are naturally related to the changes of viscous time. But then, as follows from our data, on average this time cannot be less than 11 days, and most likely, it is several tens of days.

\subsection{
Orbital modulation of matter velocity in relativistic jets}

\begin{figure}
	\includegraphics[width=\columnwidth]{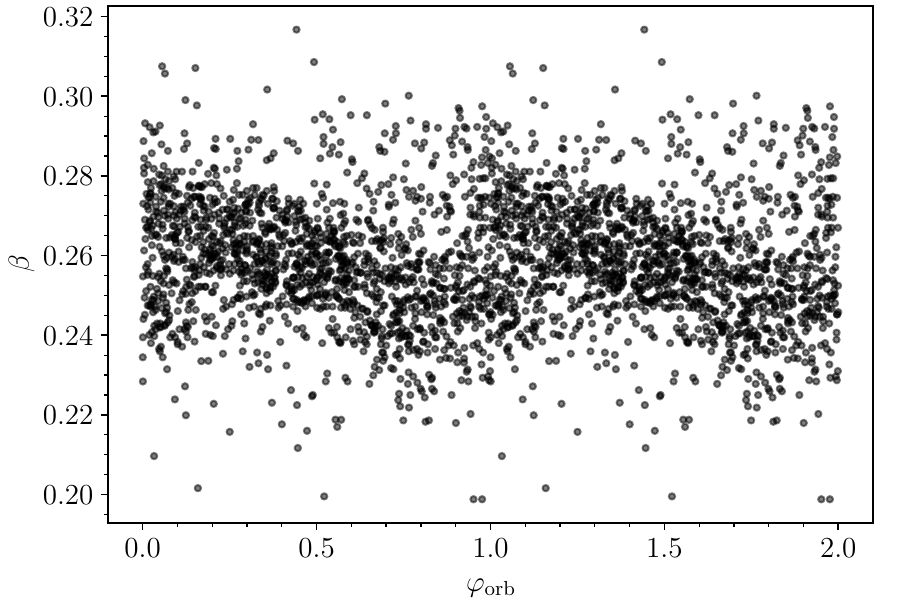}
    \caption{
    Variation of the jet velocity $\beta=v/c$ calculated from observations using formula (\ref{e:beta}) as a function of the orbital phase.}
    \label{f:beta}
\end{figure}

The new data confirm the absence of secular variations of the absolute value of matter velocity in relativistic jets. As follows from data listed in Table \ref{t:precpar}, the 45-year average value of the parameter $\beta = v/c$ is $\beta = 0.2591\pm0.0007$. The entire 45-year series of observations reveals a periodic variability of the parameter $\beta(t)$ with the  orbital period with a full amplitude of $\sim 0.04$ ($\Delta v\sim 12\,000$ km/s) -- see Fig. \ref{f:beta}. The modulation of the velocity $\beta(t)$ with the orbital period was noted in Ref.
\cite{2007A&A...474..903B,2008ARep...52..487D,2018ARep...62..747C,2022ARep...66..451C}.
In the same papers, it was hypothesized that this modulation can be caused by a small ellipticity of the SS433 orbit. In this case, the distance between the components changes with the orbital period phase, which leads to a change in the mass inflow rate from the donor star into the accretion disk and may affect the velocity of matter in the jets. The ellipticity of SS433's orbit with eccentricity $e\simeq 0.05$ was discovered in paper \cite{2021MNRAS.507L..19C}
by analyzing the 45-year average light curves of this system in the phases of maximum disk opening with respect to the observer, when the eclipsing light variations are most regular and weakly affected by the physical variability of the system.

\subsection{Stationary emission line H$\alpha$}

\begin{figure}
	\includegraphics[width=\columnwidth]{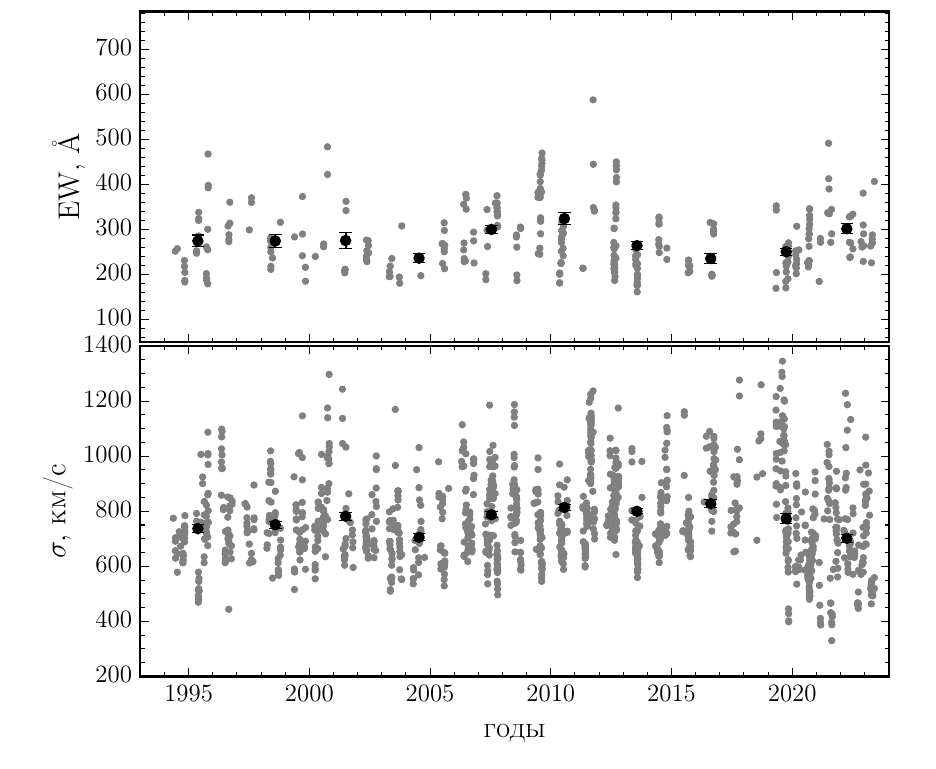}
    \caption{
    Variation of the equivalent width EW and the width $\sigma$ of the stationary H$\alpha$ line. The data for EW are given only for moments T$_3 \pm 0.2P_{\rm prec}$ and outside eclipses. Black dots are the mean values over the intervals and their errors. From paper \cite{2022ARep...66..451C} with additions.}
    \label{f:ewhalpha}
\end{figure}

\begin{figure}
	\includegraphics[width=0.75\columnwidth]{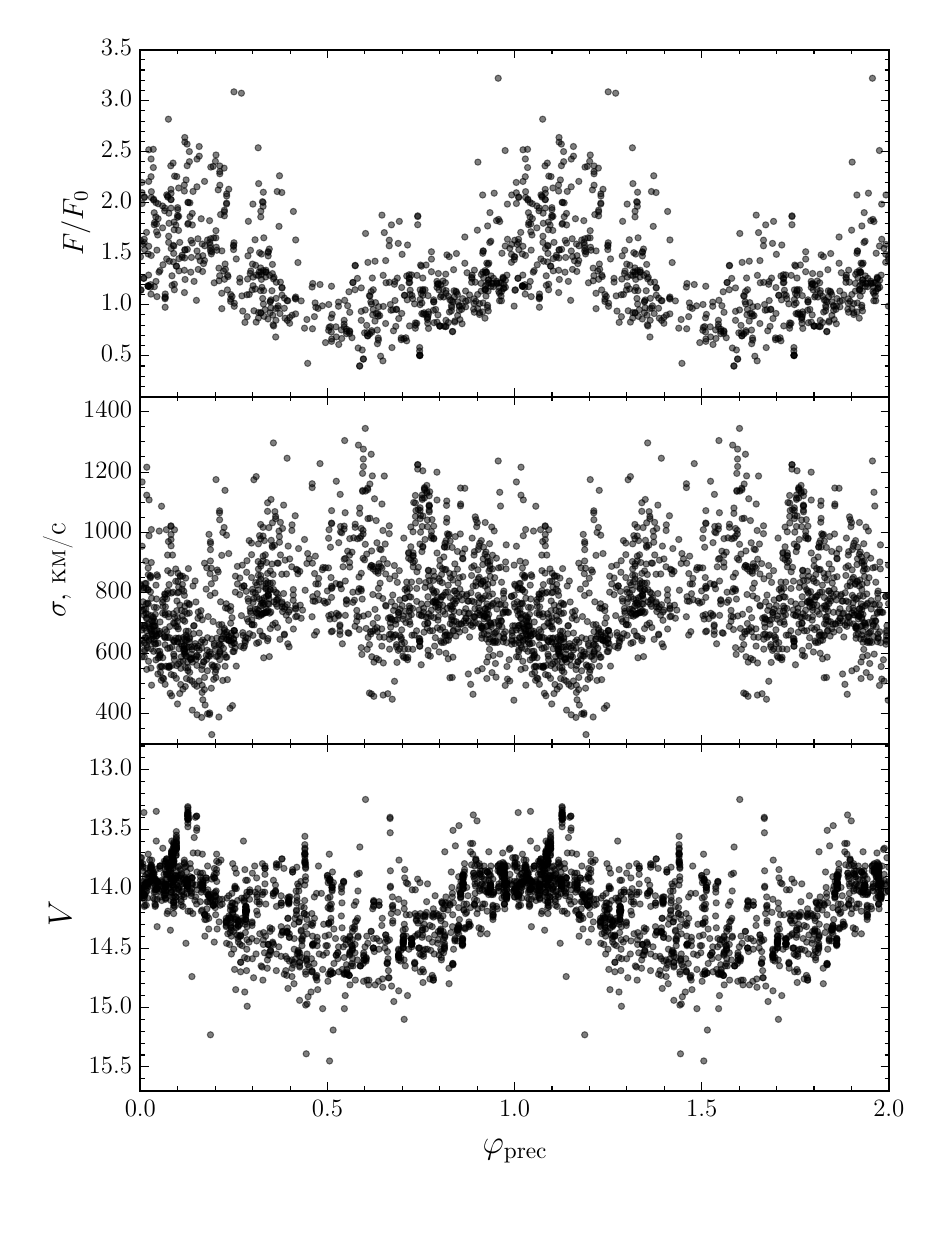}
    \caption{
    Variation of the relative flux (outside eclipses) and width $\sigma$ of the stationary H$\alpha$ line as a function of  the precessional phase.  The summary off-eclipse precessional light curve is shown in the bottom panel. From paper \cite{2022ARep...66..451C} with additions.
    }
    \label{f:halpha_prec}
\end{figure}

The stationary H$\alpha$ emission also does not demonstrate significant secular changes (see Fig. \ref{f:ewhalpha}). The equivalent width of H$\alpha$ and its Doppler width $\sigma$ are on average constant over 25 years. On short timescales significant irregular deviations from the mean values are observed. Among these irregular deviations, there is a regular component: the equivalent width of the stationary emission H$\alpha$ varies with a period close to $P_\mathrm{prec}/2$ with maxima near  phases $\varphi_\mathrm{prec}\sim 0.18$ and 0.68. The flux in the H$\alpha$ line calculated from the equivalent width using the mean precessional photometric light curve as a continuum reduces the maximum by $\varphi_\mathrm{prec}\sim0.68$, leading to a flux variability in the line with $P_\mathrm{prec}$, but with a phase shift relative to the quiet light curve in the $V$ filter of about 0.15 (see Fig. \ref{f:halpha_prec}).

To quantitatively describe the effective Doppler width of the steady-state emission H$\alpha$, we used its rms deviation in the velocity scale $V$:
$$
\sigma^2=\int(V-V_0)^2f(V)dV\,,
$$
where $f(V)$ is the normalized line profile, $V_0$ corresponds to the line center.

In the variations of the Doppler width of the stationary H$\alpha$ emission, a regular precession variability is distinguished, phased in such a way that the narrowest lines are observed at the maximum disk opening (see Fig. \ref{f:halpha_prec}). This suggests a complex structure of the wind from the precessing supercritical accretion disk. Identification of this structure and physical mechanisms of its formation is a separate task.

\section{Photometric monitoring}
\label{s:photometry}

\cite{2022ARep...66..451C}.
The photometric monitoring of SS433 was carried out in parallel with the spectral monitoring. The CAS (60-cm Zeiss-600 telescope) and CMO telescopes (RC600 automated 60-cm telescope \cite{Berdnikov2020} and 2.5-m telescope \cite{2020gbar.conf..127S}) were used. Multicolor BVRI (60-cm telescopes) and UBVRI (2.5-m telescope) CCD photometric observations of SS433 were performed. All data were processed in the standard way (dark frame subtraction, correction of sensitivity non-uniformity across the frame, etc.). SS433 brightness was determined by aperture photometry with reference to 5 comparison stars. For a description of the instrumentation and details of the processing of observations, see paper 
\cite{2022ARep...66..451C}.
\begin{figure}
	\includegraphics[width=\columnwidth]{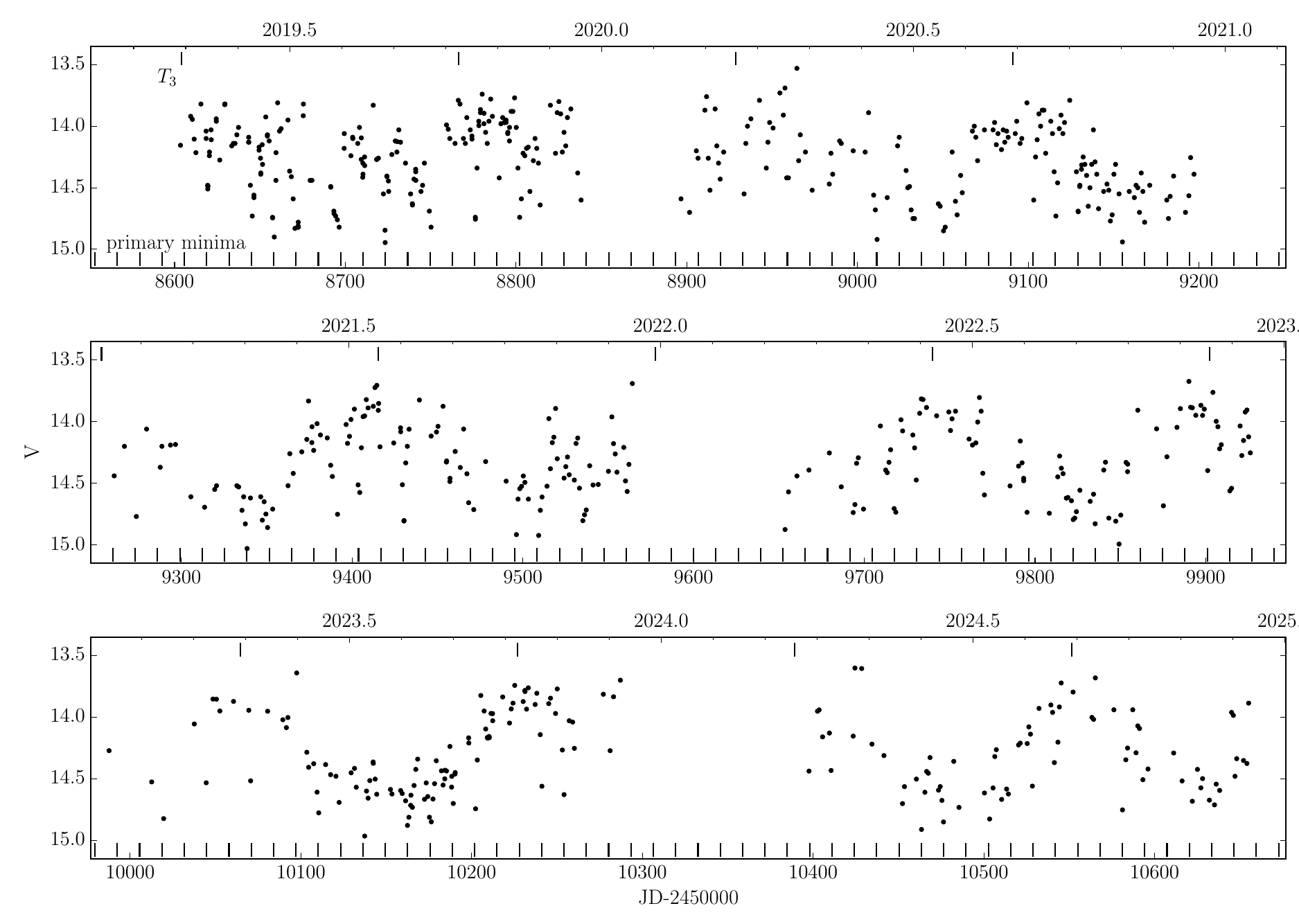}
    \caption{
   $V$ light curve from RC600 observations. Long tics at the top correspond to the moments of maximum disk opening, and tics at the bottom correspond to eclipsing minima.}
    \label{f:rc600}
\end{figure}
    
In Fig. \ref{f:rc600} we present photometric $V$-observations of SS433 obtained from 2019 to 2024. A regular precessional and eclipsing variability is visible overlaid by occasional irregular deviations and outbursts. This reflects the complex nonstationary processes accompanying the supercritical accretion regime.
    
\begin{figure}
    \centering
\includegraphics[width=\linewidth]{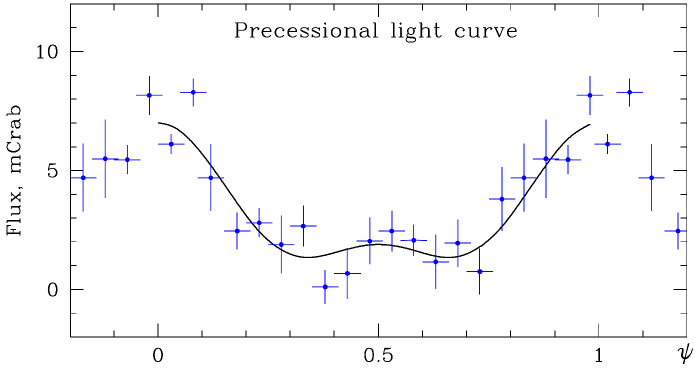}
    \caption{
    Precessional X-ray ($40-60$ keV) light curve    (blue dots). The solid curve is the model. From paper \cite{2020NewAR..8901542C}.
    }
    \label{f:precXray}
\end{figure}

Fig. \ref{f:halpha_prec} shows the convolution of all out-of-eclipse photometric observations in the $V$ filter with the precessional period. On top of the irregular variability, a regular precessional variability with an amplitude of the order of one stellar magnitude, caused by the change in the projection of the precessing accretion disk with a hot corona onto the sky plane, is clearly revealed. In the hard X-ray range $18-60$ keV, a similar precessional light curve shows a secondary maximum at phase $\varphi_\mathrm{prec}\sim0.5$
\cite{2020NewAR..8901542C}
because the relatively thin precessing accretion disk in this phase is visible from the other side (see Fig. \ref{f:precXray}).  No secondary maximum on the precessional light curve is seen in the optical. This indicates the presence in the optical of a photosphere over the disk with geometrical thickness much larger than in hard X-rays from the hot corona.

\subsection{Precessional and orbital  UBVRI light curves}

As noted above, SS433 luminosity experiences three types of regular periodicity: precessional ($P_\mathrm{prec}\approx 162^d.3$), orbital (eclipsing $P_\mathrm{orb}\approx 13^d.1$), and nutational ($P_\mathrm{nut}\approx 6^d.29$). The precessional and orbital periodicities have the largest amplitude.

For each precession phase, there is a different orbital light curve that changes with the precessional phase. However, because the precessional period is $\sim$12 times longer than the orbital one, we can neglect the change in the orbital light curve over a small interval of precessional phases (e.g., over a phase interval corresponding to one orbital period) and average the orbital light curves over small intervals of the precessional period phases.

\begin{figure}
	\includegraphics[width=0.8\columnwidth]{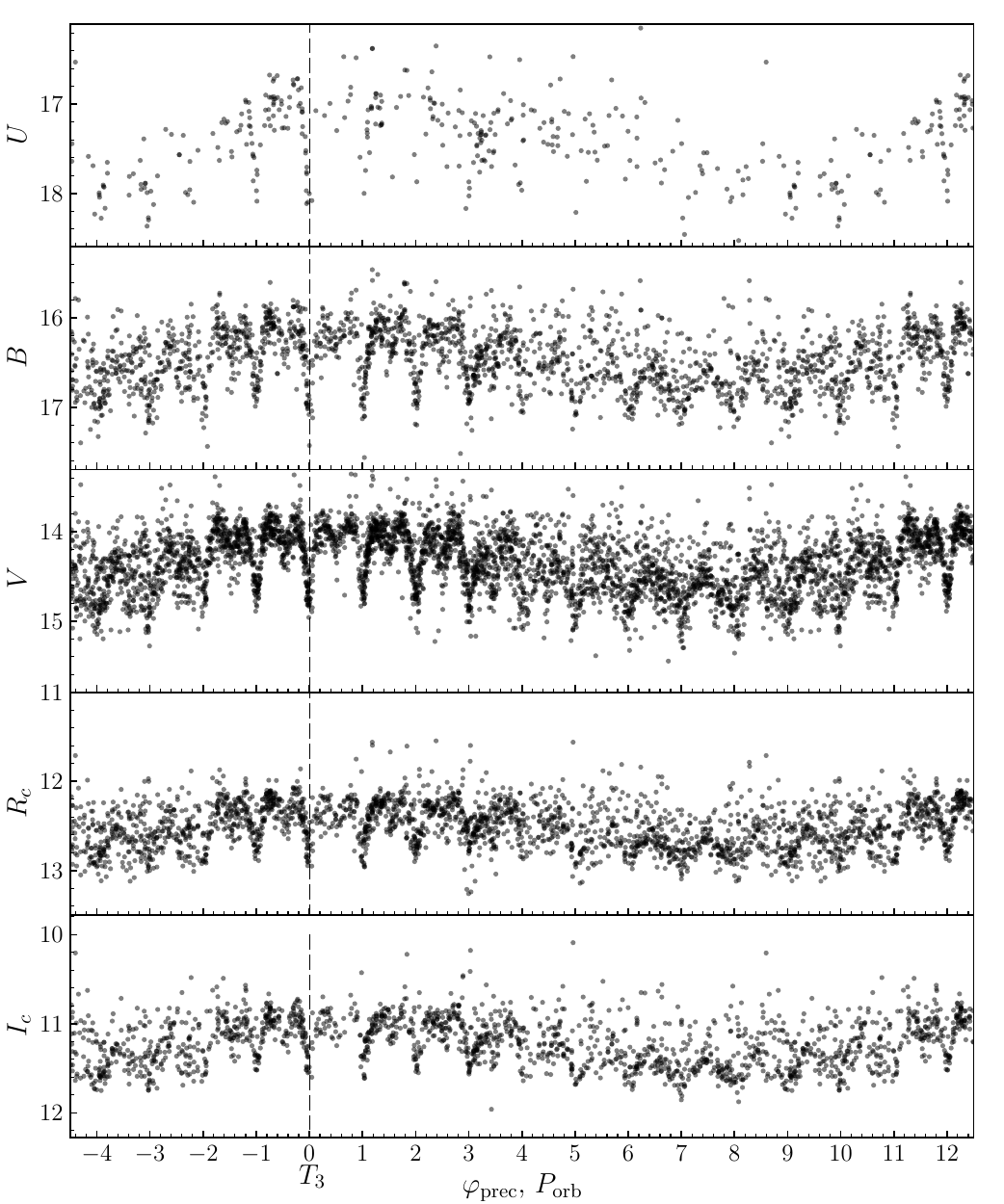}
    \caption{
    Precession-orbital variability of SS433 over all 45-year observational data. Semitransparent points are used to better display the overlapping data. Note that in the phases near the crossovers ($5 \div 9$ intervals), we observe increased physical activity of the system separating the precessional variability into two states -- high and low. Since these effects are weakly manifested in the phases of maximum disk opening (intervals $-2 \div +2$ around the moment $T_3$ marked by the vertical dashed line), the precessional variability in the high state is practically absent. From paper \cite{2024ARep...68.1349D}.}
    \label{f:prec_orb}
\end{figure}


In Fig. \ref{f:prec_orb} we show the precession-orbital $UBVRI$ light curves of SS433 constructed in this way. It can be seen that the amplitude of the precession variability as well as the depth of the eclipses grow with shortening wavelength. The regular precessional and eclipsing variability is superimposed by irregular flares. The greatest regularity of the eclipsing light curve is observed during the maximum opening of the accretion disk to the observer. In the main eclipse minimum the donor star eclipses the accretion disk, while in the secondary minimum the disk is eclipsed by the star.

\subsection{
Eclipse light curve at the maximum accretion disk opening}

\begin{figure}
	\includegraphics[width=0.75\columnwidth]{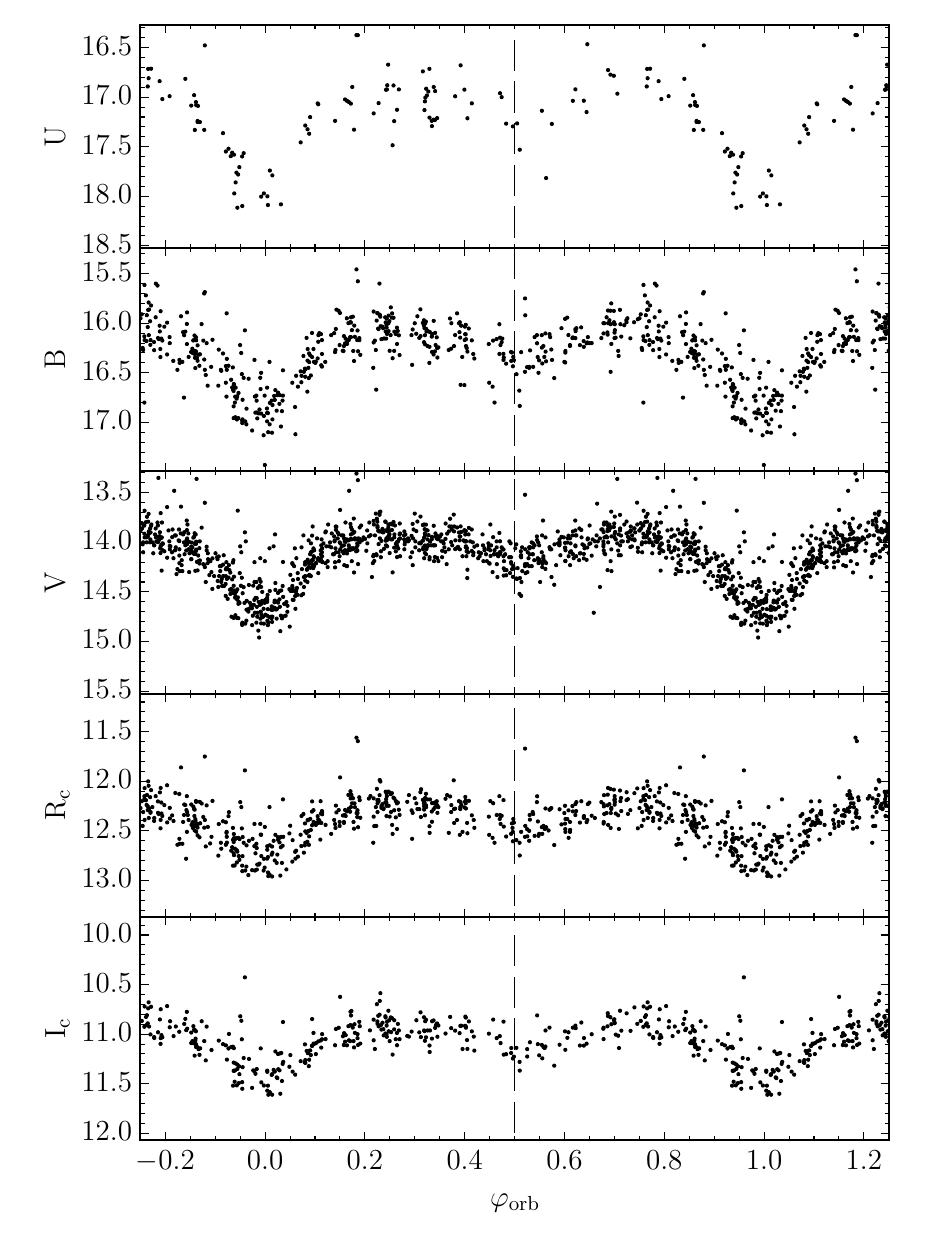}
    \caption{
    Orbital light curve near the precessional maximum $T_3\pm0.1P_{\rm prec}$ (disk is maximally open). It is seen that the secondary minimum is slightly shifted relative to the phase $\varphi_\mathrm{orb}=0.5$.
    }\label{f:orb0}
\end{figure}

\begin{figure}
	\includegraphics[width=0.75\columnwidth]{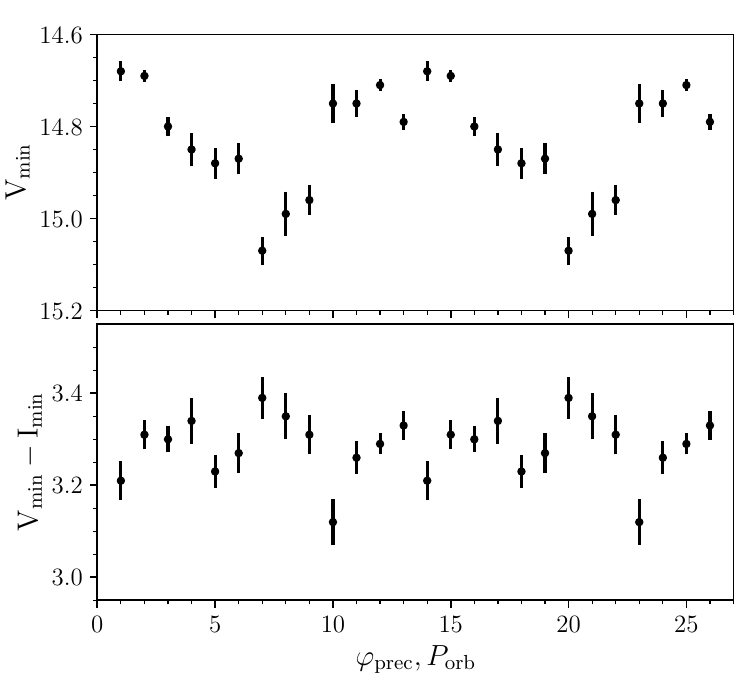}
    \caption{
    Photometric variability of minimum light and colour in the main minimum as a function of precessional phase. From paper \cite{2024ARep...68.1349D}.
    }
    \label{f:vimin}
\end{figure}

Fig. \ref{f:orb0} shows the orbital eclipsing $UBVRI$ light curves of SS433  corresponding to the maximum opening of the precessing accretion disk with respect to the observer (the moment $T_3$). The specific features of these eclipsing curves can be seen.

\subsubsection{
The third not eclipsed light in the system}

The depths of the main and secondary eclipse minima grow synchronously with wavelength shortening. This indicates the presence of a third light in the system, which is not obscured by the components and whose contribution decreases with wavelength shortening. The parameters of this third light are calculated in paper
\cite{2024ARep...68.1349D}.
In that paper we show that the main sources of the third not eclipsed light in the SS433 system with a huge wind mass-loss rate of $\sim 10^{-4}\,{\rm M}_\odot$/year are the peripheral semitransparent regions of the supercritical accretion disk, as well as the extended parts of the powerful wind from the disk. The contribution of the not eclipsed third light to the total luminosity of the system increases from 15\% in the $B$ filter ($\lambda\approx4500$\AA) to 40\% in the $I_c$ filter ($\lambda\approx 8000$\AA). The spectrum of the third not eclipsed light in the range $\lambda 4000 \div 8000$\AA\ corrected for the interstellar absorption (the total absorption in the filter $V$ is $A_V\sim 5\div 9$ magnitudes) can be described by the power law $F(\lambda) \sim \lambda^{-1}$.

\subsubsection{
Absorption in the common envelope as a cause of out of eclipse variability
}

The donor star in SS433  overflows its Roche lobe and is strongly tidally deformed. The orbital revolution of this star can lead to a significant off-eclipse variability of SS433 (ellipsoidality effect). The out-of-eclipse variability is observed in the orbital light curves (see Fig. \ref{f:orb0}). However, the contribution of the donor star to the total optical luminosity of the system is relatively small (not exceeding 26\%), and the optical emission of SS433 is dominated by the contribution of the supercritical accretion disk, from which the ellipsoidality effect should not be observed. Therefore, the out-of-eclipse variability of SS433 is mainly caused not by the ellipsoidality effect from the donor star, but by absorption of radiation from the system components in the common translucent shell during the orbital motion of the components. The common absorbing shell is formed by powerful stellar winds from the disk and the optical star.
    
At a mass loss rate of $10^{-4}\,{\rm M}_\odot$/year, the matter density between the components and in the circumbinary shell must be large enough for the effect of such absorption to be significant. This is confirmed by calculations of the light curves of eclipsing binaries with powerful wind \cite{2024ARep...68.1239A}.

\subsubsection{
The orbital ellipticity of SS433}

\begin{figure}
	\includegraphics[width=\columnwidth]{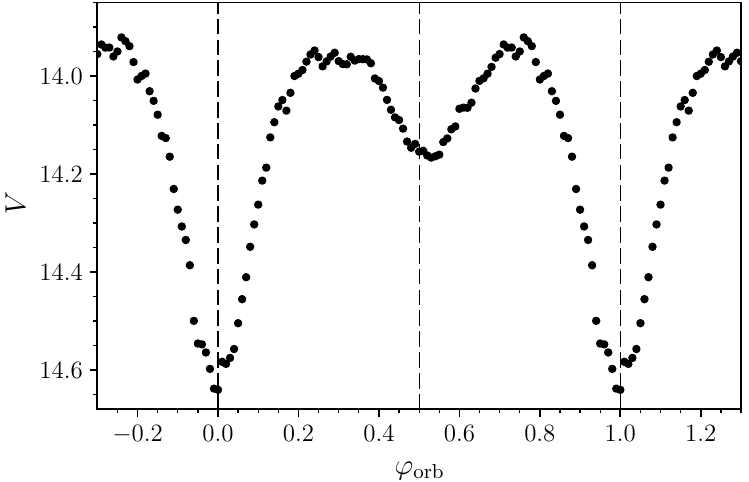}
    \caption{
     Mean orbital light curve of SS433 from data
    $1978-2012$ in the precessional phase interval $T_3\pm0.2P_\mathrm{prec}.$.
    The phase shift of the secondary minimum relative to $\varphi=0.5$ indicates the orbital eccentricity of the binary. From paper \cite{2021MNRAS.507L..19C}.
     }
    \label{f:pc}
\end{figure}
As seen from Figs. \ref{f:orb0} and \ref{f:pc}, in all $UBVRI$ eclipsing light curves the secondary minimum is shifted relative to the center between the two main minima by $\sim$0.02 orbital phase, indicating that SS433's orbit is eccentric rather than circular, as previously thought on the basis of fewer observational data. The discovery of the ellipticity of SS433 orbit with eccentricity $e\sim 0.05 \pm 0.01$ was reported in paper
\cite{2021MNRAS.507L..19C}.
    
Previously, due to the significant physical variability of SS433 and insufficient observational data, a shift of the secondary minimum in the eclipsing light curve of SS433 relative to the centre between the two main minima was not observed, so a circular orbit model was assumed for SS433. However, in the case of SS433, the circular orbit does not fit the slaved accretion disk model. The slaved disk model seems very attractive because there is a natural mechanism for it. A small asymmetry of the supernova explosion accompanying the formation of a relativistic object can turn the orbital plane of the binary system relative to the spin axis of the optical star. As a result, a binary system appears in which the rotational axis of the optical star is not perpendicular to the orbital plane. The star begins precessing under the action of tidal forces of the relativistic object, which leads to the formation of a slaved accretion disk. But if the spin axis of the optical star is not perpendicular to the orbital plane of the system, the optical star rotates non-synchronously with the orbital revolution.
\cite{1977A&A....57..383Z}
\cite{2021MNRAS.507L..19C}
The theory of synchronization of axial and orbital rotation \cite{1977A&A....57..383Z} suggests that the synchronization of the axial and orbital rotation of the binary components occurs earlier than the orbital circularization. But then, if we accept the circular orbit model for SS433, we must assume that the synchronization of axial and orbital rotation of the optical star in SS433  has had time to complete, i.e. the rotational axis of the donor star must be perpendicular to the orbital plane. In this case, the accretion disk would lie in the orbital plane and would not change its orientation relative to the observer. No precessional variability would be observed. discovery of the ellipticity of the orbit of SS433
\cite{2021MNRAS.507L..19C}
relaxes this contradiction and strongly supports the slaved accretion disk model in this system. The discovery of the orbital ellipticity of SS433 also allows us to explain variations of the matter velocity in relativistic jets of SS433 with orbital period (see above).

\subsection{
Precessional variability of minimum luminosity of SS433 in eclipse minima}

Fig.  \ref{f:prec_orb} suggests that both out-of-eclipse flux and flux at the eclipse minimum (where the accretion disk is eclipsed by the donor star) change with the precessional period of SS433. The mean light curve $V_{\rm min}$ in the V filter and the mean colour indices $V_{\rm min}-I_{\rm min}$ for the moments of the main minimum as a function of the phase of the precessional period $P_\mathrm{prec}$ are shown in Fig. \ref{f:vimin}. It can be seen that the amplitude of the regular variability of the brightness in the main minimum in the $V$ filter is $0^m.3 \div 0^m.4$, and the colour index $V_{\rm min}-I_{\rm min}$ varies almost irregularly within $\pm0^m.1$.
	
It should be emphasized that since the orbital inclination $i = 79^\circ$ of SS433 is reliably determined by the moving emissions (see Table \ref{t:precpar} above), for a small binary mass ratio  $q = M_{\rm x}/M_{\rm v} = 0.15$ found from the analysis of X-ray eclipse durations
\cite{1996PASJ...48..619K},
due to the relative smallness of the Roche lobe size of the relativistic object, a total eclipse of the accretion disk by the donor star  should be observed \cite{1987SvA....31..295A}.
In this case, the minimum flux in the middle of the main eclipse should not vary with the precessional period, which contradicts observations (see Figs \ref{f:prec_orb}, \ref{f:vimin}). A significant variability of the minimum eclipse flux with the orbital period phase  favors partial eclipses of the accretion disk by the donor star in SS433. This means that the binary mass ratio is close to unity, suggesting the presence of a black hole in SS433.
	
There is some possibility that the decrease in the minimum eclipse flux of SS433 in the phases of crossover (the disk is seen edge-on) and in the phase $\varphi_\mathrm{prec}$ = 0.5 (the disk is seen from the other side) can be due to an enhanced absorption of light from the system in an inhomogeneous circumbinary envelope. But in this case both flux and color of radiation should change. The color index $V_{\rm min}-I_{\rm min}$ for the minimum eclipse flux of SS433 with the precessional period phase does not reveal a noticeable regular component, but shows some tendency to increasing (reddening) during the eclipse, while free-bound absorption in ionized matter, on the contrary, would lead to a bluish color. The reddening tendency may be due to the fact that in these phases (disk edge-on) radiation from the colder outer parts of the disk is observed.  Thus, Fig.  \ref{f:prec_orb}, \ref{f:vimin} confirm the presence of partial eclipses in SS433 suggesting a mass ratio $q$ close to unity \cite{2024ARep...68.1349D}.

\subsection{
Eclipsing light curves in different phases of the precessional period}

\begin{figure}
	\includegraphics[width=\columnwidth]{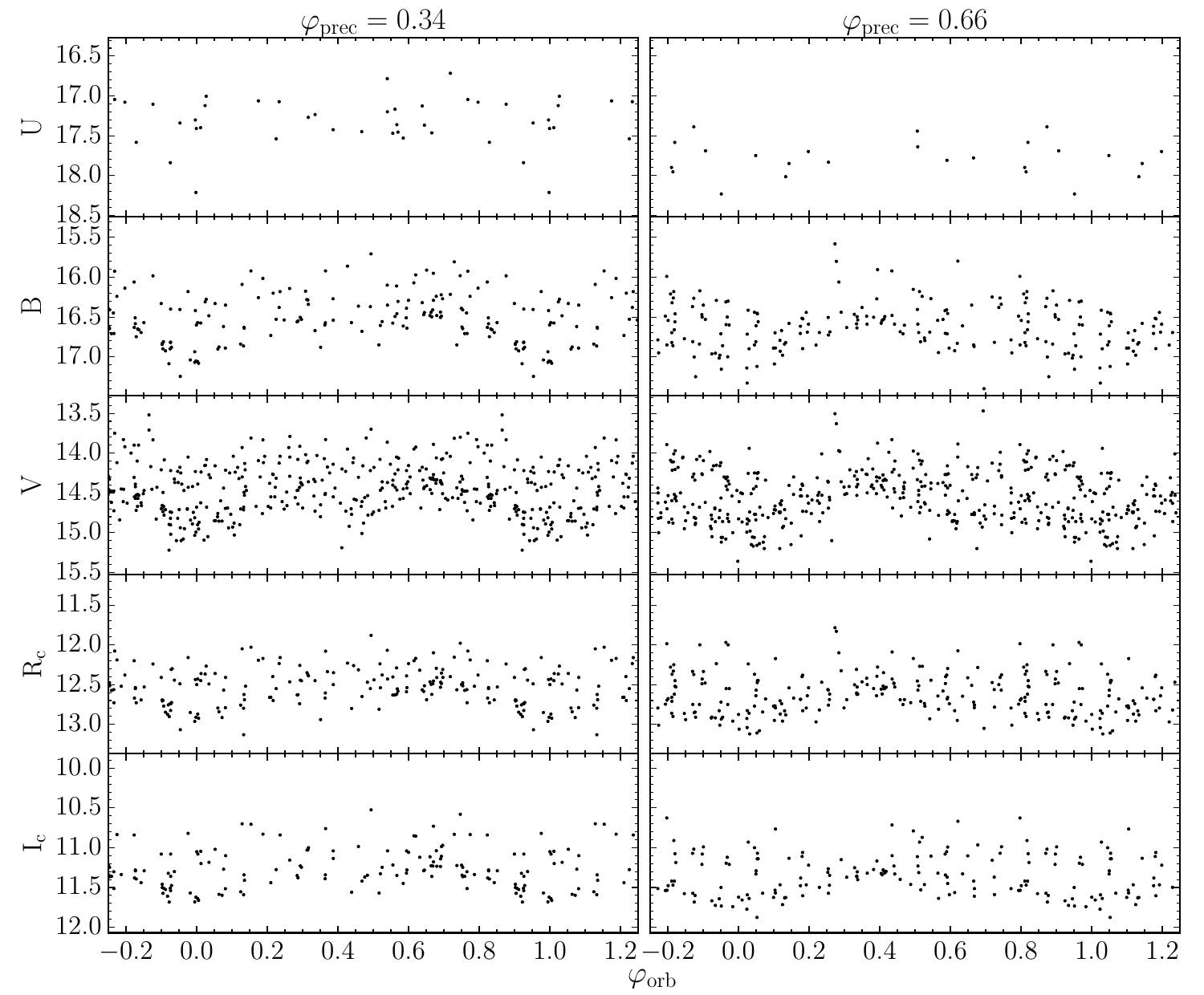}
    \caption{
    Orbital light curve near crossovers ($\varphi_\mathrm{prec}=0.34\pm0.05$, left, and $\varphi_\mathrm{prec}=0.66\pm0.05$, right). The light curve in all filters splits into two parts differing by the mean level by an almost $0^m.5$.
    }
    \label{f:orb0.34}
\end{figure}


\begin{figure}
	\includegraphics[width=0.75\columnwidth]{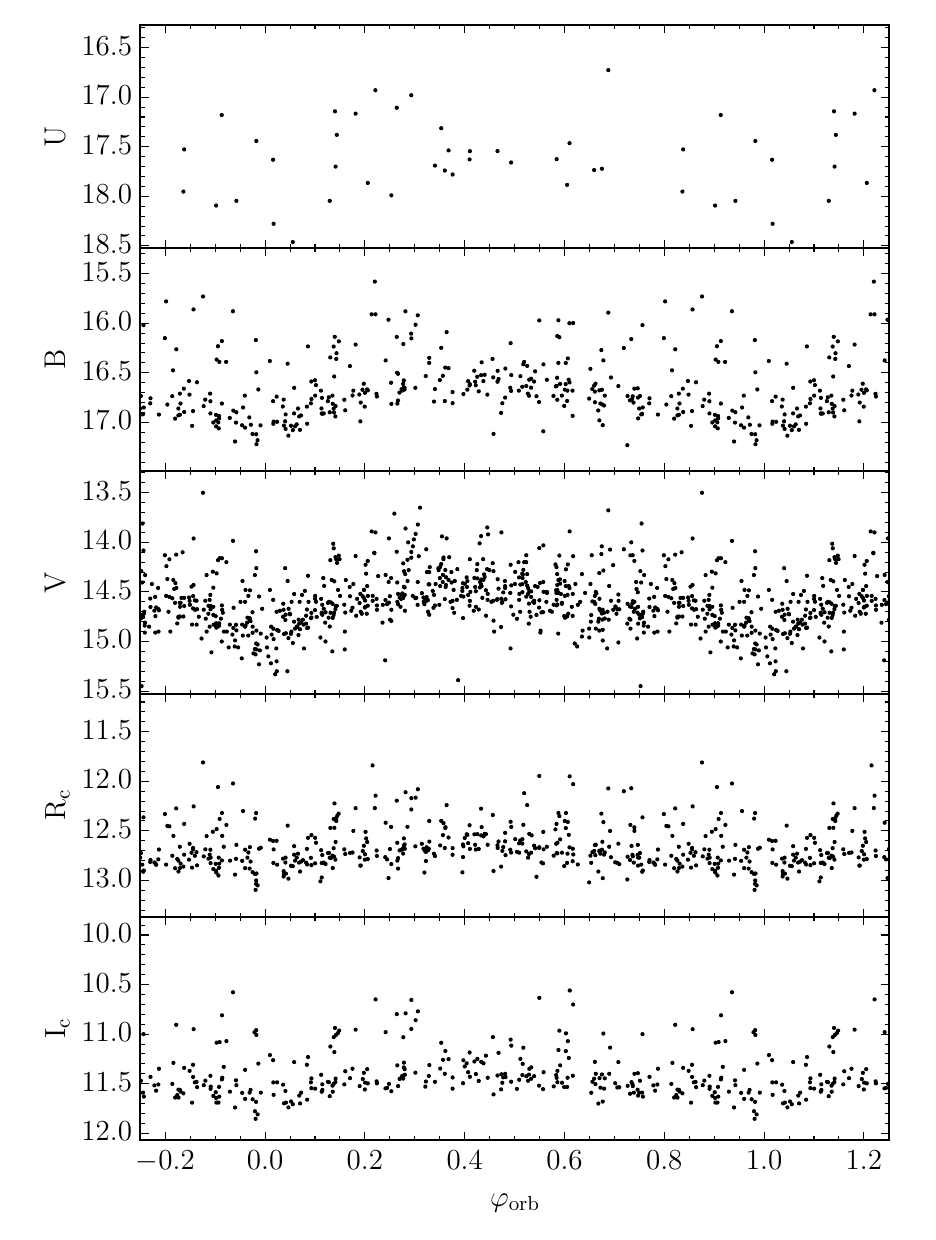}
    \caption{
    Orbital light curve near the maximum visibility of the backside of the disk $\varphi_\mathrm{prec}=0.5\pm0.1$. Two separate states as in the crossovers are not traced, but enhanced flaring activity is seen.}
    \label{f:orb0.5}
\end{figure}

In Fig. \ref{f:orb0.34}, 
\ref{f:orb0.5} we present light curves of SS433 in phases far from maximum accretion disk opening ($\varphi_{\rm prec} = 0$).

\begin{enumerate}


\item Phase $\varphi_\mathrm{prec}$ of the first crossover - the moment of the first coincidence of the positions of the moving emissions $\varphi_\mathrm{prec}\approx 0.34$ (the disk is seen edge-on).

\item Phase $\varphi_\mathrm{prec}$ when the disk is seen from the other side, $\varphi_\mathrm{prec}$ = 0.5 (the plane of the disk is tilted with respect to the line of sight by an angle $\sim 9^\circ$).

\item Phase of $\varphi_\mathrm{prec}$ of the second crossover - the moment of the second coincidence of the positions of the moving emissions $\varphi_\mathrm{prec}\approx0.66$ (the disk is seen edge-on).
\end{enumerate}



Note that the phase $\varphi_\mathrm{prec} \sim0,$ when the disk is maximally open to the observer, corresponds to the angle between the disk plane and the line of sight $\sim31^\circ$.

As seen from Fig. \ref{f:orb0.34}, 
\ref{f:orb0.5}, the shape of the SS433 light curves in phases far from the moment of the maximum disk opening  differs significantly from the light curve of a classical eclipsing binary. Clarification of the reasons for this difference is a separate problem.

That the orbital curves in the crossover phases (see Fig. \ref{f:orb0.34}) clearly reveal two states was unexpected. In the high state, the mean level of the orbital curve is increased and almost does not differ from the mean light of the system at the time of the maximum opening of the accretion disk (the precessional variability is strongly reduced). In the low state, the mean luminosity of the system is strongly reduced compared to the mean luminosity at the time of the maximum opening of the accretion disk (large amplitude of the precessional variability).
This unusual variability of SS433 in crossover phases was previously noted in \cite{2003nvm..conf..276F}.
As the precessional variability is due to the flattening of photosphere of the supercritical accretion disk, the presence of these two states in SS433 suggests changes in the disk photosphere flattening: the shape of the supercritical disk photosphere changes over a time on the order of a few days, much shorter than the orbital period.
The elucidation of the nature of the two-component structure of orbital light curves of SS433 in crossovers is a separate, as yet unsolved problem.

\subsection{
Secular evolutionary increase in orbital period}

The orbital period of SS433 $P_\mathrm{orb}\approx 13^d.1$ demonstrates remarkable stability despite the strong wind from the disk and the intense secondary mass exchange in the system on the thermal timescale of the optical star. From the approximate constancy of the orbital period of SS433 and analysis of the balance of angular momentum loss from the system,   papers \cite{2018MNRAS.479.4844C,2019MNRAS.485.2638C} inferred a large binary mass ratio $q = M_{\rm x}/M_{\rm v} > 0.6$  and concluded that there is a black hole in the system.
    

The accumulation of a large number of photometric observations of SS433, in particular in the interval 2019 - 2024 using the RC600 automated 60-cm telescope (Caucasian Mountain Observatory of SAI MSU), remotely controlled from Moscow, together with the use of all available photometric observations of SS433 since 1979 (see, e.g., the database published by V.P. Goransky), allowed us to discover a secular increase in the orbital period of this system \cite{2021MNRAS.507L..19C}.
The 45-year series of photometric observations was divided into 9 time intervals with a duration of about 4-5 years. We selected observations in the phases of the precessional period close to the maximum opening of the accretion disk, when the eclipsed light curve demonstrates the most regular shape (see Fig. \ref{f:orb0}). We then searched for change in the orbital period using two methods.

\begin{figure}
	\includegraphics[width=\columnwidth]{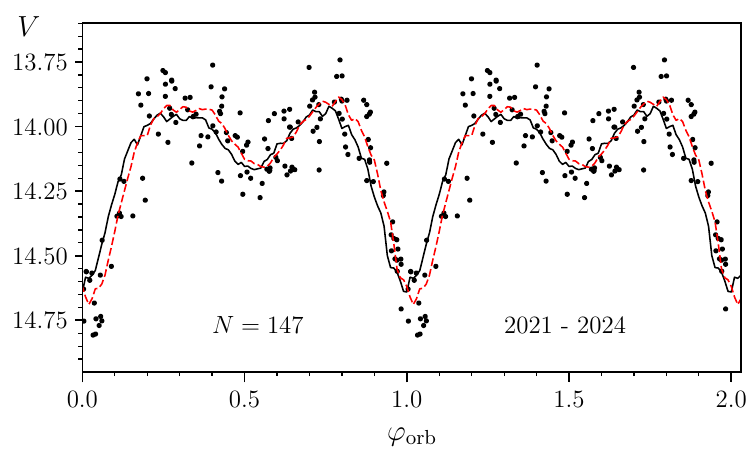}
    \caption{
    Orbital light curve of SS433 (points) near the precessional phase of the maximum disk opening $T_3\pm0.2P_{\rm prec}$ from photometry data $2021-2024$ (147 points in total). The solid black curve is the mean orbital light curve from the averaged observational data for the ephemeris with a constant orbital period. Red dashed line -- the orbital light curve phase-shifted and amplitude-scaled to the $2021-2024$ data, showing a significant shift of the minimum, which corresponds to an increase in the orbital period (Hertzsprung method).}
    \label{f:orbcurve}
\end{figure}

\begin{figure}
	\includegraphics[width=\columnwidth]{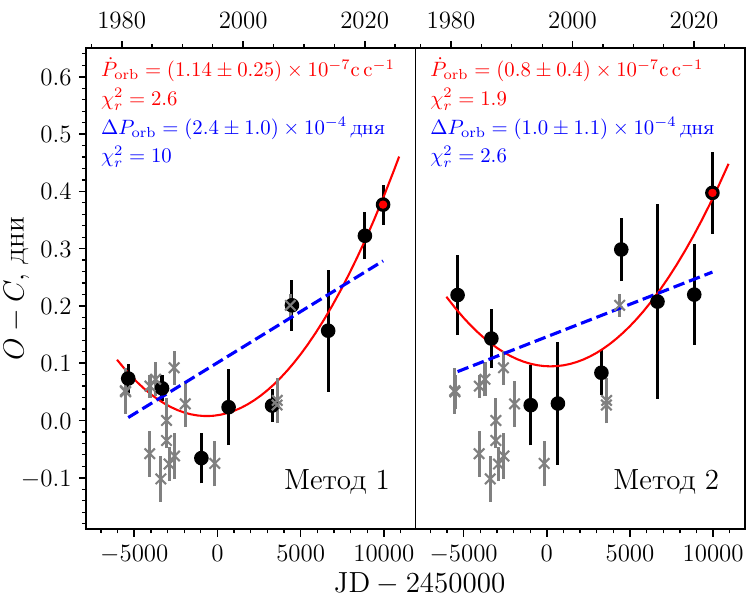}
    \caption{
 Residual deviations of the $O-C$ orbital eclipse minima of SS433 relative to the ephemeris with constant orbital period $P_\mathrm{orb}=13^\mathrm{d}.08223$ constructed by different methods. Grey crosses with error bars mark observational data from the literature. Red curve shows best-fit parabolic fit corresponding to increasing orbital period. Dotted blue line is linear approximation of the data. In the inset the time derivative of the orbital period
 $\dot{P}_\mathrm{orb}$ and a correction to the value of the mean period over the entire observation time are shown. From paper \cite{2023NewA..10302060C} with the addition of new data $2023-2024$.
    }
    \label{f:oc}
\end{figure}

    \begin{enumerate}
\item The Hertzsprung method, where the average eclipse light curve for all 45 years (black curve in Fig. \ref{f:orbcurve}) is calculated and superimposed on each individual light curve (red curve and dots in Fig. \ref{f:orbcurve}) corresponding to each time interval. The standard deviation of the individual light curve from the shifted mean curve is minimized. The resulting value is the $O-C$ plot, the time shift between the mean and the individual light curve. The left panel of Fig. \ref{f:oc} shows the time dependence of the observed $O-C$ values. We also show approximations of the $O-C$ plot using a straight line (meaning that the period does not change, but needs to be refined) and a parabola (meaning the period increases with time). It is seen that the parabolic approximation is much more preferable (the values of the minimum reduced $\chi^2_r$ differ by a factor of $\sim4$). This gives us good reason to believe that the orbital period of SS433 increases with time at a rate of ($1.14\pm 0.25)\times 10^{-7}$ seconds per second.

\item A method based on approximating the main minimum of the light curve using a Gaussian. By superimposing the Gaussians on the light curves corresponding to the selected time intervals, one can plot the $O-C$ values (see Fig. \ref{f:oc} on the right).
\end{enumerate}

Both methods lead to the conclusion that the orbital period increases, but the Hertzsprung method gives a more accurate quantitative result, from which a new ephemeris for the middle of the main minimum of the \cite{2023NewA..10302060C} is calculated:
\begin{center}
\begin{tabular}{crrl}
 $T_{\rm min}=$ &2451737.54 &  $+13.08250E$       &$+7.3\times10^{-7}E^2.$ \\
               &  $\pm$.03 &   $\pm$.00005\,\,\,\,\,        &  $\pm$1.6 
\end{tabular}
\end{center}

To explain the secular increase of the orbital period of SS433 we consider a physical model with component masses $\mv$ and $\mx$ at the non-conservative mass exchange stage under the following assumptions.

\begin{itemize}

    \item 
    The mass loss from the system is completely determined by the mass-loss rate from the optical star $\dmv$, the mass growth of the compact object is negligibly small $\dmx=0$;
    \item 
    The fraction $\beta$ of the transferred mass through the Lagrangian point ${\rm L}_1$ further outflows through the supercritical accretion disk in a completely isotropic way from the compact object as a disk wind: $\dot M_w=\beta \dmv$;
    \item 
    The fraction $(1-\beta)\dmv$ of the mass lost by the optical star leaves the binary system through the external Lagrangian point ${\rm L}_2$ and forms a disk-like shell in the orbital plane around the system with specific momentum $K$ (in units of orbital momentum).
\end{itemize}

From the total momentum conservation in this physical model of a binary system with strongly non-conservative mass exchange via a supercritical accretion disk around a compact object and additional mass loss from the system through the external Lagrangian point (see detailed derivation in \cite{2018MNRAS.479.4844C, 2019MNRAS.485.2638C,2021MNRAS.507L..19C}) we obtain the rate of change in the binary's large semi-major axis:

\beq{e:adota}
\frac{\dot a}{a}=-2\left(1-\frac{1}{2}\frac{M_\mathrm{v}}{M}\right)\frac{\dot M_\mathrm{v}}{M_\mathrm{v}}+2\beta\frac{M_\mathrm{v}}{M_\mathrm{x}}\frac{\dot M_\mathrm{v}}{M}+2(1-\beta)\frac{\left(\frac{dJ}{dt}\right)_\mathrm{out}}{J}\,,
\eeq
where $J=(M_\mathrm{x}M_\mathrm{v}/M)\sqrt{GMa}$ is the orbital angular momentum.
The first term in this formula corresponds to a fully conservative mass exchange, in which $\mv+\mx=M=const$, $\beta=0$, $K=0$. The second term is responsible for the isotropic re-emission of matter from the system under the action of radiation pressure from the supercritical accretion disk (the so-called Jeans mode), in which the specific angular momentum carried away by the wind is equal to the orbital momentum of the compact object\footnote{This approximation is valid for the supercritical disk in SS433, since the spherization radius is much smaller than the Roche lobe of the compact object, see below in Section \ref{s:model}.}. The last term is conveniently parameterized by the specific angular momentum of matter $K$ carried away from the system through the external Lagrangian point ${\rm L}_2$, in units of orbital momentum. In particular, for the mass loss through a disk of radius $R_\mathrm{out}$ in the equatorial plane of the orbit with angular velocity $v_\phi(R_\mathrm{out})$ we have
\beq{e:K}
K=\frac{v_\phi(R_\mathrm{out})R_\mathrm{out}}{(GM_\mathrm{v})^{2/3}}\myfrac{2\pi}{P_\mathrm{orb}}^{1/3}\,.
\eeq

 Substituting  $P_\mathrm{orb}=13^\mathrm{d}.1$ into formula (\ref{e:K}) and normalizing $v_\phi$ and $R_\mathrm{out}$ by $220$~km/s and 0.7 milliarcseconds obtained from VLTI GRAVITY observations of gas outflows in the vicinity of SS433 \cite{2019A&A...623A..47W}, we get
\beq{e:Knum}
K\approx 5.1 \myfrac{v_\phi(R_\mathrm{out})}{220\,\mbox{km\,s}^{-1}}\myfrac{R_\mathrm{out}}{0.7\mbox{mas}}
\myfrac{M_\mathrm{v}}{15\,{\rm M}_\odot}^{-2/3}\,.
\eeq
This is extremely large value, and the physical mechanism for the additional angular momentum supply into such a disk-like shell is unclear.

However, the observed quantity in SS433 is not the orbital semi-major axis $a$, but the orbital period $P_\mathrm{orb}$, which is related to the semi-major axis by Kepler's third law. In the case of non-conservative exchange from the system we obviously have the relation $2\dot P_\mathrm{orb}/P_\mathrm{orb}=3(\dot a/a)-\dot M/M$, from which we finally obtain the formula for the rate of change of the orbital period in our model:
\beq{e:dPP}
\frac{\dot P_\mathrm{orb}}{P_\mathrm{orb}}=-\myfrac{\dot M_\mathrm{v}}{M_\mathrm{v}}\myfrac{3q^2+2q-3\beta -3K(1-\beta)(1+q)^{5/3}}{q(1+q)}
\eeq

We note an important feature of this formula -- regardless of the mass-loss rate from the optical star, the sign of the the orbital period change will be determined only by the mass ratio $q(\beta,K)$.

Thus, with the measured rate of orbital period increase $\dpb = (1.14\pm0.25)\times 10^{-7}$\,c\,c$^{-1}$ and assumption on the ratio $\dmv/\mv<0$, equation (\ref{e:dPP}) provides constraints on the component mass ratio $q=\mx/\mv$ in the system (see \cite{2018MNRAS.479.4844C,2019MNRAS.485.2638C,2021MNRAS.507L..19C,2023NewA..10302060C} for details). The main findings of this analysis are as follows.

\begin{enumerate}
    \item 
    The component mass ratio is confidently bounded from below by the value $q\gtrsim 0.8$. With the mass of the optical star in SS433 $\mv\approx 10\,{\rm M}_\odot$ \cite{2011PZ.....31....5G}, derived from the known apparent stellar magnitude and distance to SS433 $d\approx 5$~kpc. This means that the mass of the compact object confidently exceeds the upper mass limit of neutron stars $\sim 3\,{\rm M}_\odot$, i.e., the compact object in SS433 is a black hole with mass $\mx\sim 8\,{\rm M}_\odot$. Such a mass is close to the average mass of black holes in X-ray  binaries \cite{2016PhyU...59..910C}.
    \item 
    In the framework of the physical model of non-conservative mass transfer in SS433, small values of the mass ratio $q\lesssim 0.2$ are excluded, otherwise we would observe a secular \textit{decrease} in the orbital period of the binary system, which contradicts the observations. Thus, the dynamical arguments exclude the possibility of the presence of a neutron star in SS433.
    \item  
    With an optical star mass of $10-15\\,{\rm M}_\odot$ and a flow rate $\dmv\sim 3\times 10^{-5}-10^{-4}\,{\rm M}_\odot$/yr, close to the expected mass transfer rate on the thermal timescale of the star $t_{\rm KH}=GM^2/RL$, $\dmv\sim \mv/t_{\rm KH}\approx 3\times 10^{-5}\,[{\rm M}_\odot\,\hbox{year}^{-1}] (\mv/10\,{\rm M}_\odot)^3$, the observed increase in the orbital period also implies an increase in the large semi-major axis of the binary system. The  analysis \cite{2023NewA..10302060C} shows that the parameters of the SS433 binary system are close to the case where the size of the Roche lobe remains almost constant during the mass exchange. This  can explain the fact that SS433, contrary to the theoretical predictions of the standard scheme of evolution of massive close binary systems \cite{1973NInfo..27...70T}, does not enter the stage with a common envelope, but remains a semi-detached close binary system at the stage of supercritical accretion onto a compact object.
    \end{enumerate}

It was also shown that the model of a third body in a system in which the binary orbital period can also vary is confidently rejected because the expected mass of the third body must exceed 16\,${\rm M}_\odot$. If such a massive star in the SS433 system were present as a third body, its spectral lines would be easily detected, which contradicts the spectral observations.

\subsection{Optical flares}

Fig. \ref{f:prec_orb} shows that SS433 exhibits significant irregular flaring variability on characteristic times of a few days and with an amplitude of up to 0.5 $\div $ 1 magnitude, of the same order as the amplitude of the regular precessional and eclipsing variability.
    

It is important to note that the flares are observed both in phases outside eclipses and in eclipse phases when the central part of the accretion disk is screened by the donor star. A particularly striking example of such a flare is given in a recent paper on the photometric monitoring of SS433 from the TESS space observatory \cite{2025arXiv250302753W}.
In this case, a flare of about 2-days duration was observed near the phase of the main minimum of the eclipsing light curve, with the observed optical flux increased by a factor of $\sim 3$. Studying the nature of flares in SS433 is a separate problem. The flaring activity of SS433 and identification of two photometric states of this system (active and passive) was studied earlier in paper 
\cite{2003nvm..conf..276F}.

\section{
Physical model, supercritical accretion disk and evolutionary status}
\label{s:model}

\subsection{
Supercritical accretion in SS433}

Since the pioneering work of N.I. Shakura and R.A. Syunyaev \cite{1973A&A....24..337S}, supercritical (super-Eddingtonian) accretion on compact objects (neutron stars, black holes) has been in the focus of high-energy astrophysics. This accretion mode is the physical basis for the interpretation of observations of microquasars (including SS433) \cite{1999ARA&A..37..409M,2004ASPRv..12....1F}, ultra-luminous X-ray sources \cite{2015NatPh..11..551F,2021AstBu..76....6F,2023NewAR..9601672K} and is invoked to explain the rapid mass growth of supermassive black holes in the nuclei of galaxies at large redshifts \cite{2016MNRAS.458.3047P,2024Natur.636..594J}. SS433 is the most well-studied object of this class and also has unique features -- precession and nutation of the disk and jets.

The key difference between supercritical disks and standard Keplerian disks in the theory of Shakura and Syunyaev is the presence of the so-called  spherization radius of the accretion flow, starting from which the local energy release exceeds the Eddington luminosity limit $L_\mathrm{Edd}=\frac{4\pi GM_{\rm x} c}{\chi}$ ($G$ is the gravitational constant, $M_{\rm x}$ is the accretion mass, $c$ is the speed of light, $\chi$ is the opacity of the accreting matter). In the case of a purely hydrogen plasma and Thomson scattering, $\chi=\sigma_{\rm T}/m_{\rm p}\approx 0.4$ cm$^2$/g, $m_p$ is the proton mass, and the Eddington limit is $L_\mathrm{Edd}\approx 1.3\times 10^{38} \hbox{erg/s} (M_{\rm x}/{\rm M}_\odot)$. When the local Eddington limit is exceeded, the radiation pressure begins to sweep matter out with a parabolic velocity from radius $r_{\rm s}$, determined from the relation $\frac{1}{2}\frac{GM_{\rm x}\dot M}{r_{\rm s}}=L_\mathrm{Edd}$, where $\dot M$ is the accretion rate in the disk.
Thus, in the supercritical regime only a small fraction of matter accretion rate corresponding to $\dot M(r)=\dot M_{\rm cr}(r)$ penetrates inside the sphericalization radius and reaches the last stable orbit around the black hole, which defines the inner boundary of the disk. Often the accretion rate is normalized to a critical value corresponding to the Eddington luminosity of accretion onto a black hole $ L_\mathrm{Edd}\approx 0.1\dot M_{\rm cr} c^2$, $\dot M_{\rm cr}(M_{\rm x})\approx 10^{-7}\,[{\rm M}_\odot/\hbox{year}](M_{\rm x}/10\,{\rm M}_\odot)$.

For SS433, $\dot M\sim 10^{-4}\,{\rm M}_\odot\hbox{/year}=10^3\dot M_{\rm cr}(10\,{\rm M}_\odot)$. The sphericalization radius is $r_{\rm s}\approx 7.4\times 10^9\,[\hbox{cm} ](\frac{M_{\rm x}}{10\,{\rm M}_\odot})(\frac{\dot M}{1000\,\dot M_{\rm cr}})$. The characteristic outflow velocity of matter from the sphericalization radius is of the order of the parabolic velocity $v_{out}=\sqrt{\frac{2GM_{\rm x}}{r_{\rm s}}}$ $=c\sqrt{0.4 {\frac{\dot M_{\rm cr}}{\dot M}}}\simeq 6000\,\frac{\hbox{km}}{\hbox{s}}\sqrt{\frac{1000\,\dot M_{\rm cr}}{\dot M}}$.
At the strongly supercritical accretion rates $\dot M\gg \dot M_{\rm cr}$ characteristic of SS433, an opaque photosphere is formed in the strong wind from the disk with a characteristic radius $r_{\rm ph}\sim r_g(M_{\rm x})(\dot M/\dot M_{\rm cr})^{3/2}\sim 10^{11} \hbox{cm}(\frac{\dot M}{1000\,\dot M_{\rm cr}})^{3/2}(\frac{M_{\rm x}}{10\,{\rm M}_\odot})$ \cite{2007MNRAS.377.1187P} ($r_g=2GM_{\rm x}/c^2\approx 30\,[\hbox{km}](M_{\rm x}/10\,{\rm M}_\odot)$ is the gravitational radius of the accretor).
The realistic radius of the photosphere above the disk of SS433 estimated  from the analysis of the observed luminosity and temperature is an order of magnitude higher, $r_{\rm ph}\sim 10^{12}$~cm \cite{1986ApJ...308..152W}, i.e., comparable to the size of the Roche lobe of the compact object in SS433 $R_{\rm L}(M_{\rm x})\sim 2\times 10^{12}$~cm. This difference is due to the non-sphericity of the outflowing matter and a funnel-like shape of the photosphere confirmed by numerical multidimensional radiative MHD calculations (see, for example, the recent work \cite{2024MNRAS.532.4826T} with the binary system parameters close to those of SS433). The main characteristic scales of the supercritical disk in SS433 are given in Table \ref{t:super}.


\begin{table}
\caption{Characteristic dimensions of the SS433 binary system and supercritical accretion disk}
 \label{t:super}
\begin{tabular}{lcc}
\hline
\hline
Parameter & Value for $q=1$ and $M_{\rm x}=10{\rm M}_\odot$\\
\hline
Major semi-axis of the orbit & $\approx 7\times 10^{12}$ cm\\
Roche lobe radius $R_{\rm L}(M_{\rm x})$ & $\approx 2.6\times 10^{12} $ cm\\
Spherization radius & $\sim 7.4\times 10^9$ cm\\
Radius of photosphere &$\sim 2.5 \times 10^{12}$ cm\\
\hline
\end{tabular}
\end{table}

\begin{figure}
    \centering
\includegraphics[width=0.5\linewidth]{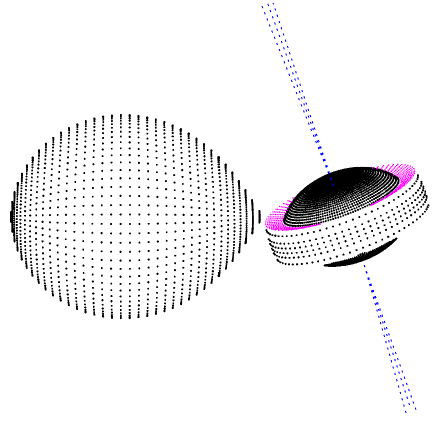}
    \caption{
    Schematic model of microquasar SS433 with a supercritical accretion disk, a wide accretion funnel filled with hot gas (corona), and relativistic jets (from X-ray data).}
    \label{fig:scheme}
\end{figure}

Numerical calculations in general confirm the expected picture of supercritical accretion on a compact object -- a powerful outflow of matter under the action of radiation pressure from the inner parts of the accretion disk and the formation of a photosphere in the wind, as well as the presence of an accretion funnel in the wind with an opening angle of about 90$^\circ$ filled with high-speed rarefied hot gas (see Fig. \ref{fig:scheme}). It is interesting to note that exactly this geometrical model (a thick disk with a wide funnel filled with coronal hot gas) was derived from the interpretation of the SS433 eclipsing and precessional variability in hard X-rays \cite{2020NewAR..8901542C} (see Fig. \ref{f:precXray}).
We also note that, according to numerical multidimensional calculations \cite{2024MNRAS.532.4826T}, the main wind outflow from the supercritical disk occurs in the equatorial region, and the flow in the disk is convectively unstable. The characteristic turnover time of a convective cell in the accretion disk at radius $R$ is of the order of $t\sim 1/\omega_{\rm K}(R) \mathrm{Ri}^{-1}$, where Ri$<1$ is the Richardson number \cite{2010MNRAS.404L..64L}. At the outer radius of the accretion disk in SS433 of the order of the Roche lobe of the compact object $R_{\rm out}\sim 2\times 10^{12}$ cm, the Keplerian time is $t_K\sim 1/\omega_{\rm K}(R_{\rm out})\approx 21$ hours.
Therefore, in the supercritical accretion flow, variability associated with convective instability of the disk can be expected on a timescale of $t_{\rm var}\lesssim 20^h/\mathrm{Ri}\sim$ several days. Note that it is in this time interval that quasi-periodic optical oscillations (QPOs) with a characteristic period of $\sim 13$ hours are sometimes observed \cite{2025arXiv250302753W}. This model is supported by the fact that on the precessional-orbital curve shown in Fig. \ref{f:prec_orb} the amplitude of flaring activity is maximal in the red $I_{\rm c}$ filter and minimal in the violet $U$ filter.

\subsection{
Features of accretion disk and optical star precession in SS433}

As seen from Fig.  \ref{f:OCprec}, the precessional variability keeps stable over large time intervals, which indicates the presence of a ''setting clock mechanism''. For SS433, this is thought to be the precession of an optical star with the rotational axis tilted to the orbital plane, which leads to the formation in the neighborhood of the inner Lagrangian point ${\rm L}_1$ of a gas stream flowing out of the orbital plane and an inclined precessing disk. The inclined disk precesses under the combined action of tidal forces and the dynamical action of the gas stream (a similar precessing disk is observed in Her X-1, see \cite{1999A&A...348..917S} for details). In this model, the phase jumps of the precessional variability, clearly visible in Fig. \ref{f:OCprec}, occur due to the change in the dynamic influence of the gas stream on the outer parts of the disk during inevitable variability of the gas flow rate through the inner Lagrangian point, the change in the inclination angle of the disk outer parts, etc.
The maximum amplitude of the precessional period jumps should not exceed the characteristic viscous time in the accretion disk, which for supercritical accretion in SS433 should be of the order of an orbital period: $t_{\rm visc}\sim P_\mathrm{orb}(R/H)^2/\alpha$, where $H/R\sim 1$ for supercritical disks, $\alpha<1$ is the dimensionless Shakura-Sunyaev viscosity parameter \cite{1973A&A....24..337S}. Obviously, the maximum amplitude of the observed precession period jumps in SS433 ($\pm 20$ days) falls inside this interval.

As shown in paper \cite{1999A&A...348..917S} (see formula (2) in that paper), the period of tidal motion of the outer parts of the disk with radius $R_{\rm d}$ and inclination $\theta$ to the orbital plane $P_\mathrm{tid}\sim P_\mathrm{orb} R_{\rm d}^{-3/2}/\cos\theta$. For the SS433 parameters ($\theta=19^\circ$, component mass ratio $q=M_{\rm x}/M_{\rm v}=0.8$, $R_{\rm d}/a\approx 0.5$), we obtain the ratio of the purely tidal precessional period of the disk to the orbital period $P_\mathrm{tid}/P_\mathrm{orb}\sim 10$, while the observed value is $P_\mathrm{prec}/P_\mathrm{orb}\sim 13$.
This suggests, as in the case of Her X-1, the braking dynamical action of gas streams interacting with the outer parts of the disk and creating a moment of forces opposite to the tidal one. Note that a variation of the outer disk radius will lead to a change in tidal precession period by the amount $\Delta P_\mathrm{tid}/P_\mathrm{tid}\sim -3/2 \Delta R_{\rm d}/R_{\rm d}$. Clearly, a 10\% variation of the disk radius is sufficient to produce beats between the stable precession period of the star and the disk with a characteristic time of the order of $5-7$ precessional periods. This is the relaxation time  after phase jumps to the mean value seen in Fig. \ref{f:OCprec}.

In the SS433 system, precession of the optical star tilted to the orbital angular momentum by the angle  $\theta\approx 20^\circ$ is assumed. The main contribution to the precession of an inclined rotating star in a binary system of the SS433 type will come from the tidal interaction of the rotating star with the second component (relativistic effects are small). In this case the angular frequency of precession of the star's angular momentum vector (spin) for a circular orbit is equal to
\cite{1975PhRvD..12..329B,1985SvAL...11..224S}
\[
\frac{\Omega_{\rm prec}}{\omega_\mathrm{orb}}=-\frac{\bar\omega}{\omega_{\rm orb}}\frac{k_2}{\eta}q\left(\frac{R}{a}\right)^3\sin\theta\cos\theta
\]
where $k_2$
is the apsidal motion constant, $\eta$ is the dimensionless coefficient in the star's moment of inertia $I=\eta M R^2$, $\bar\omega$ is the characteristic angular velocity in the star's angular momentum, $J=I\bar\omega$, $\omega_\mathrm{orb}=2\pi/P_\mathrm{orb}$ is the orbital angular frequency, $q=M_{\rm x}/M_{\rm v}\le 1$ is the binary mass ratio. The minus sign means precessional motion in the direction opposite to the orbital revolution. The precessional to orbital period ratio is
\[
\frac{P_\mathrm{prec}}{P_\mathrm{orb}}=\frac{\omega_\mathrm{orb}}{\bar\omega}\frac{\eta}{k_2}\frac{1}{q}\left(\frac{a}{R}\right)^3\frac{1}{\sin\theta\cos\theta}
\]
At $q\sim 0.8$ and full synchronization of the star's rotation with the orbital motion, we obtain $P_\mathrm{prec}/P_\mathrm{orb}\sim8\eta/k_2\sim 200-300$ for a characteristic $k_2\sim of 10^{-2}$ and $\eta\sim 0.25$, while the observed period ratio is $P_\mathrm{prec}/P_\mathrm{orb}\sim 13$. This suggests a faster rotation of the optical star (more precisely, its inner parts) compared to the orbital frequency $\bar\omega/\omega_\mathrm{orb}\sim 10$. This situation seems quite possible for SS433, as the age of the W50 nebula left after the supernova explosion is on the order of $10-100$ thousand years, during which the star does not have time to arrive at a tidally synchronized state with orbital motion.
Before the supernova explosion during the first mass exchange, the optical star in SS433 could have been spun up by accretion to the limiting Keplerian velocity of about 400 km/s (such rotational velocities are typical for fast-rotating Be-stars with masses up to $\sim 10\,{\rm M}_\odot$ in hard X-ray transient binaries). The present-day rotational velocity of the outer layers of the optical star in SS433 (based on the broadening of the spectral lines) is about 140 km/s 
\cite{2020A&A...640A..96P}, which is close to the orbital velocity. This means that the outer layers of the star have had time to synchronize on the time interval of the order of $10^4-10^5$ years.
However, the large ratio $\bar\omega/\omega_\mathrm{orb}\sim 10$ required to explain the tidal precession of an inclined optical star with a period of 162 days requires a faster rotation of the star (mainly of its inner layers). For a short time ($\sim 10^4-10^5$ years) after the supernova explosion and the formation of the relativistic object in SS433, a complete synchronization of the orbital rotation and axial rotation of the entire star might had not occur. This is indirectly supported by the presence of a small orbital eccentricity $e\approx 0.05$ in SS433 \cite{2021MNRAS.507L..19C}, as well as the inclination of the optical star's rotation axis to the orbital plane ($\sim 19^\circ$).

\section{Hard radiation from SS433}
\label{s:hard}

Microquasar SS433 has been and continues to be observed in hard X-rays and gamma-rays from space-based X-ray and gamma-ray observatories and ground-based gamma-ray observatories, in particular, RXTE
\cite{2006A&A...460..125F,2015MNRAS.446..893A,2016AstL...42..517A}, XMM-Newton \cite{2010MNRAS.402..479M,2018AstL...44..390M}, 
Chandra \cite{2003PASJ...55..281N,2013ApJ...775...75M,2019AstL...45..299M}, NuSTAR\cite{2021MNRAS.506.1045M,2023A&A...669A.149F}. 
From the X-ray observations of SS433, the temperature of the radiative thermal plasma at the base of relativistic jets is calculated to be $kT_0\sim 12-18$ keV (depending on the precession phase), the total luminosity of the jets and disk is $L_{\rm x}(2-70\,\hbox{ keV})=2-20\times 10^{37}$\,erg/s, and the opening angle of jets is $\Theta<2^\circ$. A high-resolution X-ray spectroscopy of SS433 \cite{2013ApJ...775...75M} suggests a significant (an order of magnitude higher than solar) nickel abundance in the jet emission. The increased nickel abundance in the disk wind (although by a factor 1.5 less than in the jets) was derived from the analysis of the XMM-Newton spectra \cite{2018AstL...44...390M}. This indicates an unusual chemistry of an optical star contaminated with heavy elements in a supernova explosion after which a compact object was formed.

Very successful were long-term observations of SS433 X-ray eclipses in the hard X-ray range $18-60$ keV from the INTEGRAL space gamma-ray observatory carried out in 2013-2011 in different phases of the precessional cycle with a total duration of $\sim 8.5$ Ms \cite{2020NewAR..8901542C}, which allowed us to construct the orbital and precessional light curves of the source mentioned above. The main results of the analysis of hard X-ray observations of SS433 are as follows.
\begin{enumerate}
 \item 
  For the first time non-thermal hard X-ray emission with a photon index of $\Gamma\approx 3.8$ was detected, indicating the presence of a hot coronal plasma with $kT\simeq 20$ keV in the funnel of a supercritical accretion disk, which comptonizes the thermal X-ray emission from relativistic jets
 \cite{2003A&A...411L.441C,2009MNRAS.394.1674K}.
    \item 
     The width of the X-ray eclipse, the large amplitude of the orbital and precessional variability, and the independence of the shape of the hard X-ray emission spectrum from the precession phase indicate that a hot extended disk corona (see Fig. \ref{fig:scheme}) was eclipsed by the optical star, enabling us to impose independent constraints on the component mass ratio $q>0.3$ from the geometrical model of the source \cite{2009MNRAS.397..479C}.
    \item 
    From the shape of the $18-60$ keV orbital light curve and Swift/BAT X-ray monitoring data we found evidence of tidal nutation of a tilted precessing disk with a period of 6.29 days \cite{2013MNRAS.436.2004C}.
\end{enumerate}

In recent years, new observations of SS433 and the W50 nebula in the hard X-ray and gamma-ray bands appeared. Let us note some important results and problems.

\begin{enumerate}
    \item 
    Gamma-rays ($100-300$ GeV) from the SS433/W50 region are detected by the Fermi/LAT space telescope \cite{2015ApJ...807L...8B}. Ultrahigh-energy gamma-ray photons up to energies of 25 TeV are detected by the HAWC facility (up to 25 TeV) \cite{2018Natur.562...82A}. A joint analysis of the SS433 Fermi-LAT and HAWC \cite{2020ApJ...904..188K} gamma-ray observations shows their common origin in the region of jets interacting with the putative supernova remnant W50. The hard spectrum can be described in the model of Compton back-scattering of X-ray synchrophotons on electrons accelerated in the jets-shell collision region to energies $\sim 100$ TeV (lepton mechanism). The hadronic mechanism, in which X-ray emission is produced by protons accelerated to energies of the order of 23 PeV \cite{2020ApJ...904..188K}, is not excluded.

    \item 
The analysis of 10-year observations of SS433 with the Fermi-LAT gamma-ray telescope \cite{2020NatAs...4.1177L} revealed a precessional periodicity. This possibly indicates the role of a strong precessional equatorial outflow of matter from the central source in the transfer of kinetic energy to the accelerated protons. Traces of such an equatorial outflow were also obtained from the analysis of hard X-ray data from the NuSTAR \cite{2021MNRAS.506.1045M}. A powerful equatorial outflow during strongly supercritical accretion in double systems is obtained in modern numerical calculations \cite{2024MNRAS.532.4826T}.

    \item 
        The LHAASO facility  has detected gamma-ray photons in the energy range $25-100$ TeV from SS433/W50 \cite{2024arXiv241008988L}. The purely lepton model of Compton back-scattering on electrons accelerated to energies up to 200 TeV does not describe the hardest photons in the spectrum at energies above 10 TeV because of the reduced Klein-Nishina cross section of the photon-electron interaction. This supports the need to add the hadronic origin of the hard gamma-ray emission from SS433 and requires the injection of protons at energies up to 3 TeV with a power of at least $10^{38}$ erg/s. These observations suggest that Galactic microquasars and ultraluminous X-ray sources with supercritical accretion onto compact objects could be Galactic PeVatrons and explain the observed cosmic ray spectrum up to $10^{15}$ eV \cite{2024A&A...688A...4C,2024arXiv241108762P}.
    
\end{enumerate}

\section{Conclusion}
\label{s:conclusion}

Microquasar SS433 is the first example of the phenomenon of supercritical accretion on a stellar-mass black hole. Supercritical accretion was predicted in \cite{1973A&A....24..337S}. In the case of supermassive black holes, the phenomenon of supercritical accretion is especially pronounced in quasars -- very active nuclei of galaxies, where, as well as in SS433, relativistic jets are often formed. A unique feature of SS433 is that the supercritical accretion disk here is inclined to the orbital plane and, together with relativistic jets, precesses with a period of $\sim 162.3$ days. It is the presence of moving spectral emissions formed in the outer parts of precessing relativistic jets, attracted special attention to this unique object. For more than 40 years of research scientists have managed to make significant progress in understanding the nature of this object. However, a number of fundamental questions related to SS433 remained unsolved until recently. In the Introduction, we have listed the most important of them.

In recent years, thanks to the continuous spectral and photometric monitoring of SS433 performed at SAI MSU over the last 30 years, and using all published data since 1979, new progress has been made in understanding the nature of SS433. This has brought us closer to solving some questions listed in the Introduction. In particular, the ellipticity of SS433's orbit has been discovered, which provides strong support for the slaved accretion disk model that tracks the precession of the rotational axis of the donor star. From the analysis of a 46-year series of photometric observations of SS433, a secular evolutionary increase in the orbital period of this binary system was discovered. Using the angular momentum loss balance from SS433, we were able to give an independent estimate of the component mass ratio $q = M_{\rm x}/M_{\rm v} > 0.8$ and the black hole mass $M_{\rm x} \gtrsim 8\,{\rm M}_\odot$. From the same analysis it follows that during the evolution of SS433 the distance between the components of this system grows with time, which prevents the formation of a common envelope in the system. The size of the Roche lobe of the donor star in SS433 is on average constant in time, which ensures the stability of the secondary mass exchange \cite{2023NewA..10302060C}.

The parameters of the kinematic model of the SS433 binary system, except for the precessional period, are on average constant in time over a period of 45 years, which is also favors the slaved accretion disk model. We detected jumps of the precession period phase, which are apparently caused by viscous time variations in the supercritical accretion disk.

The Doppler width of the stationary  H$\alpha$ emission reaches minimum at times close to the moment of the maximum opening of the accretion disk to the observer, which points to a complex structure of the wind outflow from the supercritical accretion disk. We also detected a narrow variable absorption in the blue wing of the stationary H$\alpha$ emission, indicating nonstationary processes in supercritical accretion disk wind \cite{2025arXiv250301698D}.

The structure of the supercritical accretion disk around the black hole, as well as the structure and kinematics of the powerful wind from the disk remain unsolved issues. It is necessary to understand the reasons for the formation of relativistic jets -- whether they are caused by the supercritical accretion regime or they are formed by some special, specific mechanisms. The shape of orbital light curves at the crossovers of moving emissions in the spectrum, when the disk is visible edge-on, and the presence in these phases of the precession period of two states of the mean system light need a separate study both by observational means and with the help of theoretical modeling. Also a separate problem is the clarification of the physical mechanisms of the appearance of optical flares from SS433. Object SS433 and nebula W50 are sources of hard X-ray and gamma-ray radiation.

Object SS433 is physically close to a new class of unique objects, the ultraluminous X-ray sources discovered in recent years in many galaxies \cite{2004ASPRv..12....1F,2015NatPh..11..551F}. The ULX class presents many examples of supercritical accretion on both a black hole and a neutron star with a strong magnetic field \cite{2014Natur.514..202B}. The study of microquasar SS433 as a representative of the ULX class is of great interest for relativistic astrophysics.

Hard X-ray and TeV gamma-ray observations of SS433 and the surrounding W50 nebula suggest the possibility of hadron acceleration to ultrahigh energies of $\sim$ PeV range when a powerful outflowing equatorial and polar wind from the supercritical accretion disk in SS433 interacts with the matter of the W50 shell and makes the SS433 microquasar also a Galactic PeVatron.

Further multi-wavelength observations of this unique source will certainly make a significant contribution to the understanding of the most energetic processes of supercritical accretion onto black holes, important for solving one of the key modern astrophysical problems of the mass growth of central supermassive black holes in the nuclei of galaxies and quasars.

\section*{Acknowledgments}

The work of A.M. Cherepashchuk and A.V. Dodin was supported by RNF grant 23-12-00092 (problem formulation, observations, their processing, and participation in the interpretation of the results). The work of K.A.Postnov (theoretical interpretation of the observational results) was supported by grant 075-15-2024-541 of the Ministry of Education and Science of the Russian Federation under the program for financing large scientific projects of the national project ''Science''.


\end{document}